\documentclass[useAMS,usenatbib,usegraphicx]{mn2e}
\usepackage{aas_macros}
\usepackage[british]{babel}
\usepackage{multirow}
\usepackage{color}
\bibpunct{(}{)}{,}{a}{}{,} % to follow the A&A style
\newcommand{\lt}{<}
\newcommand{\gt}{>}
\newcommand{\cmd}{CMD}
\newcommand{\pms}{pre-MS}
\newcommand{\zams}{ZAMS}
\newcommand{\ms}{MS}

\newcommand{\std}{STD}
\newcommand{\acc}{ACC}
\newcommand{\msun}{M$_{\sun}$}

\newcommand{\lsun}{L$_{\sun}$}
\newcommand{\mearth}{M$_{\earth}$}

\title[Effect of planet ingestion on low-mass stars]{Effect of planet ingestion on low-mass stars evolution: the case of 2MASS~J08095427--4721419 star in the Gamma Velorum cluster}
\author[E. Tognelli, P. G. Prada Moroni, \& S. Degl'Innocenti]
{E. Tognelli$^{1,2}$\thanks{e-mail: emanuele.tognelli$@$yahoo.it}, P.G. Prada Moroni$^{1,2}$\thanks{e-mail: pier.giorgio.prada.moroni$@$unipi.it}, 
S. Degl'Innocenti$^{1,2}$\\
$^{1}$University of Pisa, Department of Physics 'E.Fermi', Largo Bruno Pontecorvo 3, I-56127, Pisa, Italy\\
$^{2}$INFN, Section of Pisa, Largo Bruno Pontecorvo 3, I-56127, Pisa, Italy
}
\begin{document}
\date{Accepted 2016 May 25. Received 2016 May 25; in original form 2015 December 21}
\pagerange{\pageref{firstpage}--\pageref{lastpage}} \pubyear{2016}
\maketitle
\label{firstpage}
%%
%% Abstract %
\begin{abstract}
We analysed the effects of planet ingestion on the characteristics of a \pms{} star similar to the Gamma Velorum cluster member  2MASS~J08095427--4721419 ($\#52$). We discussed the effects of changing the age $t_0$ at which the accretion episode occurs, the mass of the ingested planet and its chemical composition. We showed that the mass of the ingested planet required to explain the current [Fe/H]$^{\#52}$ increases by decreasing the age $t_0$ and/or by decreasing the Iron content of the accreted matter. 

We compared the predictions of a simplified accretion method -- where only the variation of the surface chemical composition is considered -- with that of a full accretion model that properly accounts for the modification of the stellar structure. We showed that the two approaches result in different convective envelope extension which can vary up to 10~percent. We discussed the impact of the planet ingestion on a stellar model in the colour-magnitude diagram, showing that a maximum shift of about 0.06~dex in the colour and 0.07~dex in magnitude are expected and that such variations persist even much later the accretion episode. We also analysed the systematic bias in the stellar mass and age inferred by using a grid of standard non accreting models to recover the characteristics of an accreting star. We found that standard non accreting models can safely be adopted for mass estimate, as the bias is $\la 6$ percent, while much more caution should be used for age estimate where the differences can reach about 60 percent.

\end{abstract}
\maketitle
%%
%% Sections
\section{Introduction}
\label{sec:intro}

Recently, \citet{spina14} found a chemically peculiar star, namely 2MASS~J08095427--4721419 (hereafter $\#52$), in the young Gamma Velorum cluster. This star is significantly metal richer ([Fe/H]$^{\#52}=+0.07\pm0.07$) than the cluster itself ([Fe/H]$^\rmn{cl.}=-0.057\pm0.018$), showing also an over-abundance of refractory elements (e.g. Mg, Al, Sc, Ti, V and Co), which are expected to condensate in a proto-planetary disc \citep{spina15}. \citet{spina14,spina15} suggested that those chemical abundance peculiarities might be ascribed to one or more episodes of planet engulfment, a scenario that has already been proposed to explain chemical anomalies found in Iron-rich and/or Lithium-rich hosting planets stars \citep[see e.g.,][ and references therein]{gonzalez97,laughlin97,israelian01,montalban02,ashwell05,carlberg10}. 

\citet{spina14} estimated a mass of about 1.3~\msun{} for $\#52$ relying on standard stellar models that do not take into account mass accretion episodes. At the age of the cluster, i.e. 10--20~Myr \citep[][]{jeffries14}, a star of 1.3~\msun{} is in an advanced enough stage of the \pms{} evolution to have already developed a sizeable radiative core and to have an external convective region thinner than a few percent of the total mass. 

The ingestion of a rocky planet on a star with these characteristics is thus able to produce  significant surface chemical abundance anomalies as the metal rich material is diluted in a thin convective envelope. Clearly, the extent of the anomalies depends on the planet characteristics -- its mass and chemical composition -- and on the mass of the stellar convective envelope. 

\citet{spina14} assumed that the planet accretion episode does not affect the structure of the star and in particular the extension of its convective envelope. Within such a simplified assumption, the only effect of the planet ingestion is the modification of the surface chemical composition, which can be easily computed. Following this approach, they found that the ingestion of a rocky planet with mass in the range of 50--70~\mearth{} in the last 5~Myr would be enough to produce the observed high-[Fe/H]$^{\#52}$ value.

The main aim of the present paper is twofold: 1) to extend the analysis performed by \citet{spina14,spina15} by computing stellar models that consistently follow the accretion episode and 2) to quantify the resulting differences. Unfortunately, the details of the accretion are not known and cannot be directly obtained from the observations. As such, this work is intended to be a first investigation of the main effects of a single episode of planet ingestion on a star similar to $\#52$. We focused on the impact on the results of the age at which the accretion begins and of the properties of the accreted planet, such as its mass and chemical composition. The only constraint imposed to the models is to reproduce the observed high-[Fe/H]$^{\#52}$ value after the accretion process. 

The paper is structured as it follows. In Sect.~\ref{sec:models} and Sect.~\ref{sec:acc_models} we summarized the main characteristics of the computed stellar evolutionary models with and without accretion. In Sect.~\ref{sec:feh} and~\ref{sec:heavy} we analysed the impact of the accretion on the surface chemical composition ([Fe/H], total metallicity $Z$ and heavy element abundances), discussing the differences between the full and the simplified accretion model. In Sect.~\ref{sec:cmd} we analysed the effect of the planet ingestion on the theoretical predictions for the position of the star in the Colour-Magnitude Diagram (CMD). In Sect.~\ref{sec:rec}, we discussed the systematic error induced in recovering the properties of the star that ingested a planet (i.e. mass and age) with standard evolutionary tracks that do not take into account the accretion episode.

\section{Stellar models}
\label{sec:models}
We computed the stellar models by means of the Pisa evolutionary code \citep[\textsc{prosecco},][]{tognelli15a,tognelli15b} -- an updated version of the \textsc{franec} code \citep[see e.g.][]{deglinnocenti08,tognelli11,dellomodarme12} -- adopting the same input physics/parameters and the thin-disc mass-accretion treatment described in \citet{tognelli15c}. 

To compute stellar models to be compared against stars with an observed [Fe/H] value, proper values of the initial helium and metal abundances must be chosen. To do this, we used eqs. (1) and (2) given in \citet{gennaro10} assuming the \citet{asplund09} solar-scaled metals distribution. The [Fe/H] values (hereafter [Fe/H]$_\rmn{G07}$) provided by \citet{spina14} for the Gamma Velorum cluster and $\#52$ are referred to the \citet{grevesse07} solar Iron abundance, i.e. $\log \epsilon_{\rmn{Fe},\sun}^\rmn{G07} \equiv 12 + \log (N_\rmn{Fe}/N_\rmn{H})_{\sun}= 7.45$. Such a value is slightly different from the more recent \citet{asplund09} one ($\log \epsilon_{\rmn{Fe},\sun}^\rmn{AS09} = 7.50$) adopted in our stellar models. Thus, for consistency, we converted the [Fe/H]$_\rmn{G07}$ into the [Fe/H]$_\rmn{AS09}$ using the [Fe/H] definition, namely
\begin{equation}
[\rmn{Fe/H}] \equiv \log \bigg(\frac{N_\rmn{Fe}}{N_\rmn{H}}\bigg)_\star - \log \epsilon_{\rmn{Fe},\sun} + 12
\label{eq:feh_def}
\end{equation}
In our case, applying eq.~(\ref{eq:feh_def}) we obtained, 
\begin{eqnarray}
[\rmn{Fe/H}]_\rmn{AS09} &=& [\rmn{Fe/H}]_\rmn{G07} + \log \epsilon_{\rmn{Fe},\sun}^\rmn{G07} - \log \epsilon_{\rmn{Fe},\sun}^\rmn{AS09}\nonumber\\
&=&[\rmn{Fe/H}]_\rmn{G07} - 0.05
\end{eqnarray}
By adopting, respectively, [Fe/H]$_\rmn{G07}^\rmn{cl.}=-0.057$ and [Fe/H]$_\rmn{G07}^{\#52}=+0.07$ for the bulk of the cluster and for $\#52$ \citep{spina14}, the conversion leads to [Fe/H]$_\rmn{AS09}^\rmn{cl.}=-0.107$ and to [Fe/H]$_\rmn{AS09}^{\#52}=+0.02$, which, in turn, translate into the following initial helium and metal contents, ($Y^\rmn{ini}=0.269$, $Z^\rmn{ini}=0.0102$)$_\rmn{cl.}$ and ($Y^\rmn{ini}=0.275$, $Z^\rmn{ini}=0.0135$)$_{\#52}$. 

To analyse the impact of the ingestion of a planet on stellar evolution, we firstly computed a grid of standard models (hereafter \std{} models, reference set) in the mass range [1.0,~1.4]~\msun{} with a mass spacing of $0.01$~\msun{}, evolved from the beginning of the Hayashi track -- without any accretion episode -- to the exhaustion of the central Hydrogen. We selected this mass range as the mass we obtained for the $\#52$ star is approximatively 1.2~\msun{} (see Sect. \ref{sec:simple}).

Standard non accreting models have been computed with the low-[Fe/H] value ([Fe/H]$_\rmn{AS09}^\rmn{cl.}=-0.107$, $Y^\rmn{ini}=0.269$ and $Z^\rmn{ini}=0.0102$), representative of both the bulk of the cluster and of the initial chemical composition of $\#52$, if the accretion hypothesis is fulfilled. A set of standard non-accreting models with the high-[Fe/H] value measured for $\#52$ ([Fe/H]$_\rmn{AS09}^{\#52}=+0.02$, $Y^\rmn{ini}=0.275$ and $Z^\rmn{ini}=0.0135$) is also computed, to take into account the possibility  that $\#52$ was born with a chemical composition different from the cluster one and equal to that currently observed.

As a second step, we computed models with the low-[Fe/H] value that follow the accretion episode. The planet ingestion is supposed to occur in a single episode at a stellar age $t_0$. To check the effect of changing the age at which the accretion episode occurs, we adopted three $t_0$ values, namely $t_0=10$, 12, and 15~Myr, where the last one approximatively corresponds to the estimated age of the Gamma Velorum cluster. We did not investigate cases for $t_0 \lt 10$~Myr because, as we will show in Sect.~\ref{sec:simple}, the mass of the planet required to reproduce the observed [Fe/H]$_\rmn{AS09}^{\#52}$ would reach unrealistic large values.
%%
%% Tab: Zi/Z:
\begin{table}
\centering
\caption{Metals relative abundances $\tilde{z}_i^\rmn{p}$ used for the accreted matter$^{(a)}$.}
\label{tab:zi_z}
\begin{tabular}{l|r|r||l|r|r}
\multicolumn{6}{c}{$\tilde{z}_i^\rmn{p}\equiv X_i^\rmn{p}/Z^\rmn{p}$}\\
\hline
  elem. & AS09 & JK10 & elem. & AS09 & JK10 \\
\hline
\hline
 C & 0.17699 & 0.0 & Cl & 0.00061 & 0.0 \\
 N & 0.05185 & 0.0 & Ar & 0.00549 & 0.0 \\
 O & 0.42902 & 0.31096 &  K & 0.00022 & 0.0 \\
Ne & 0.09403 & 0.0 & Ca & 0.00480 & 0.00863 \\
Na & 0.00219 & 0.0 & Ti & 0.00023 & 0.00050 \\
Mg & 0.05297 & 0.13040 & Cr & 0.00124 & 0.00361 \\
Al & 0.00416 & 0.00863 & Mn & 0.00081 & 0.0 \\
Si & 0.04976 & 0.18958 & Fe & 0.09669 & 0.32902 \\
 P & 0.00043 & 0.0 & Ni & 0.00533 & 0.01867 \\
 S & 0.02314 & 0.0 \\
\hline
\end{tabular}
\medskip
\flushleft
$^{(a)}$ The $\tilde{z}_i^\rmn{p}$ mass abundances have been re-normalised to sum to unity.
\end{table}

Since we are interested in the impact on the stellar evolution of a planet ingestion episode, with particular attention to the surface chemical composition, we analysed the effect of accreting matter with two different metal distributions. For simplicity, the accreted matter is assumed to be Hydrogen- and Helium-free, thus we adopted $Z^\rmn{p}=1$ for the metal abundance of the ingested planet. In one case we assumed the same metal mass fraction relative abundances of the star \citep[i.e. the Solar-like mixture by][hereafter AS09]{asplund09}\footnote{Metal mass fraction relative abundance is the ratio between the mass fraction abundance of the $i$-th metal $X_i$ and the total metallicity $Z$. For a Solar-like metal distribution (or mixture) $(X_i/Z)_\star = (X_i/Z)_{\sun}$.}, thus matter mainly composed by Oxygen, Carbon, Neon and Iron. In the following we will refer to this case as the Iron-poor distribution. In the other case we adopted the same metal distribution of the Earth \citep[][hereafter JK10]{javoy10}, thus a material mainly constituted by Iron, Oxygen, Silicon and Magnesium (hereafter Iron-rich case). This last case corresponds to accrete matter where the volatile elements are absent and where the condensation processes have occurred -- as expected in the case of rocky planets -- and hence compatible with the over-abundance of refractory elements measured by \citet{spina15} in the spectra of $\#52$.

The relative mass abundances of the metals $\tilde{z}_i^\rmn{p}\equiv X_i^\rmn{p}/Z^\rmn{p}$ for the two adopted distributions are listed in Table \ref{tab:zi_z}. It is important to notice that the Iron content in the two mixtures is different, being the Iron abundance in Earth-like one (JK10) about 3.4 times larger than that in the Solar-like mixture (AS09). In the next Sections, we will discuss the impact of such a difference on the models.

In the following we will show the evolutionary tracks in the observed Color-Magnitude Diagram (\cmd). To convert the models from the theoretical ($\log T_\rmn{eff}$, $\log L/$\lsun) to the observed ($B-V$, $M_V$) plane, we used the bolometric corrections computed from the synthetic spectra by \citet{allard11} -- for several $\log g$, $\log T_\rmn{eff}$ and [Fe/H] values -- and applying the formalism described in \citet{girardi02}. Since the surface [Fe/H] value of the accreting stellar models changes in time, the color transformations have been performed using the actual surface [Fe/H] at each stellar age.

\section{Accretion models}
\label{sec:acc_models}
To quantify the impact of a planet ingestion episode on stellar surface Iron abundance, we followed two different approaches, one quite simplified but with the advantage of being easily implemented by means of the commonly available standard models and the other more accurate and realistic but computationally more demanding. In the former case, similarly to \citet{spina14}, we assumed that the planet ingestion does not affect the evolution of the stellar structure, in particular, leaving unaltered the extension of the convective envelope with respect to that of standard non-accreting models. On the contrary, in the latter case, we actually followed the accretion process computing consistently the structure evolution of non-standard stellar models in presence of matter accretion. The differences between such approaches are discussed in the following sections.

\subsection{Simplified accretion models}
\label{sec:simple}

In this first type of models, following an approach similar to that of \citet{spina14}, we assumed that the only effect of accretion is the change of the surface chemical abundances. In such a scenario, the final surface Iron abundance can be easily computed assuming that the accreted metal-rich matter is completely mixed and diluted with that within the convective envelope of the standard model (i.e. non-accreting model). Clearly, for a given planet, the final surface abundance strongly depends on the dilution process efficiency, i.e. on the extension of the convective envelope, which in turn depends on the characteristics of the stellar model, i.e mass, age and initial chemical composition. 

Assuming that the matter of the ingested planet $M_\rmn{p}$ is instantaneously and fully mixed with that contained in the convective envelope $\Delta M_\rmn{c.e.}$, the  mass fraction abundance of the $i$th-element inside the convective envelope after the accretion process $X_i^\rmn{new}$ is given by the following equation:
\begin{equation}
X_i^\rmn{new} = \frac{X_i \Delta M_\rmn{c.e.} + X_i^\rmn{p}M_\rmn{p}}{\Delta M_\rmn{c.e.} + M_\rmn{p}}
\label{eq:acc}
\end{equation}
where $X_i$ and $X_i^\rmn{p}$ are the $i\rmn{th}$-element mass fraction abundance of the matter inside the stellar convective envelope before the accretion and in the ingested planet, respectively. From eq.(\ref{eq:acc}) one can immediately obtain the surface total metallicity $Z^\rmn{new}$ after the accretion episode, 
\begin{equation}
Z^\rmn{new}= \frac{Z \Delta M_\rmn{c.e.} + Z^\rmn{p}M_\rmn{p}}{\Delta M_\rmn{c.e.} + M_\rmn{p}}
\label{eq:zsup}
\end{equation}
and the corresponding [Fe/H] surface value,
\begin{equation}
[\rmn{Fe/H}]^\rmn{new} = \log \Big(\frac{X_\rmn{Fe}^\rmn{new}}{X_\rmn{H}^\rmn{new}}\frac{m_\rmn{H}}{m_\rmn{Fe}}\Big) - \log \epsilon_{\rmn{Fe},\sun} + 12
\label{eq:feh}
\end{equation} 
where $m_\rmn{H}$ and $m_\rmn{Fe}$ are the Hydrogen and Iron atomic masses and $\log \epsilon_{\rmn{Fe},\sun} = \log \epsilon_{\rmn{Fe},\sun}^\rmn{AS09} = 7.50$ \citep{asplund09}. By inverting eqs. (\ref{eq:acc}) and (\ref{eq:feh}) the mass of the ingested planet needed to obtain the requested [Fe/H]$^\rmn{new}$ can also be computed, 
\begin{equation}
\frac{M_\rmn{p}}{\Delta M_\rmn{c.e.}} = \frac{X_\rmn{H}\times 10^{[\rmn{Fe/H}]^\rmn{new}+\beta} - X_\rmn{Fe}}{X_\rmn{Fe}^\rmn{p}}
\label{eq:mp} 
\end{equation}
where $\beta=\log \epsilon_{\rmn{Fe},\sun}^\rmn{AS09} - 12 + \log m_\rmn{Fe}/m_\rmn{H}$.

It is evident from eqs.~(\ref{eq:acc})--(\ref{eq:feh}) that, once the stellar model has been fixed (i.e. mass and chemical composition), the variation of the stellar surface chemical abundances depends on $\Delta M_\rmn{c.e.}$ -- which changes with the age and hence $t_0$ -- on the ingested planet mass $M_\rmn{p}$ and on its chemical composition $X_i^\rmn{p}$. This implies that the mass of the ingested planet required to reproduce the observed surface chemical anomalies (the high-[Fe/H] value) depends only on $t_0$ and on the planet chemical composition, in particular on $X_\rmn{Fe}^\rmn{p}$ (see eq. \ref{eq:mp}), once the stellar model has been fixed.

Concerning the characteristics of the star, we adopted an initial mass $M~=~1.2$~\msun{}, as obtained to reproduce the position in the \cmd{} of $\#52$, i.e. $(B-V)=0.65$ and $M_V=3.95$ \citep{spina14}, with our own evolutionary tracks with the same metallicity of the cluster [Fe/H]$^\rmn{cl}_\rmn{AS09}=-0.107$ (i.e. $Y^\rmn{ini}=0.269$, $Z^\rmn{ini}=0.0102$) at an age of about 15~Myr (see Sect.~\ref{sec:cmd}), the mid value of the cluster age range (10--20~Myr) estimated by \citet{jeffries14}.
%% Fig: [Fe/H], Mp:
\begin{figure}
\centering
\includegraphics[width=\columnwidth]{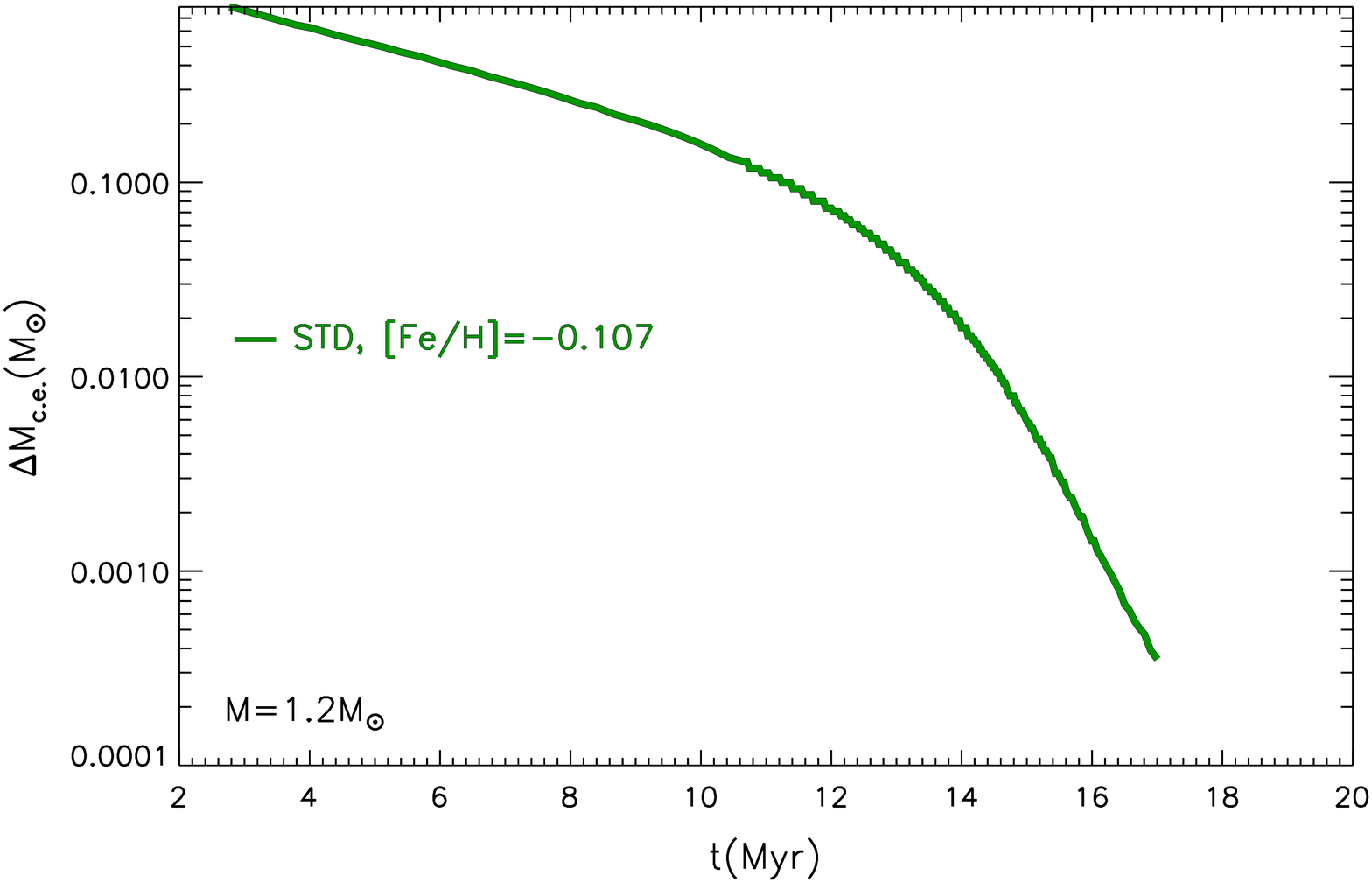}\\
\includegraphics[width=\columnwidth]{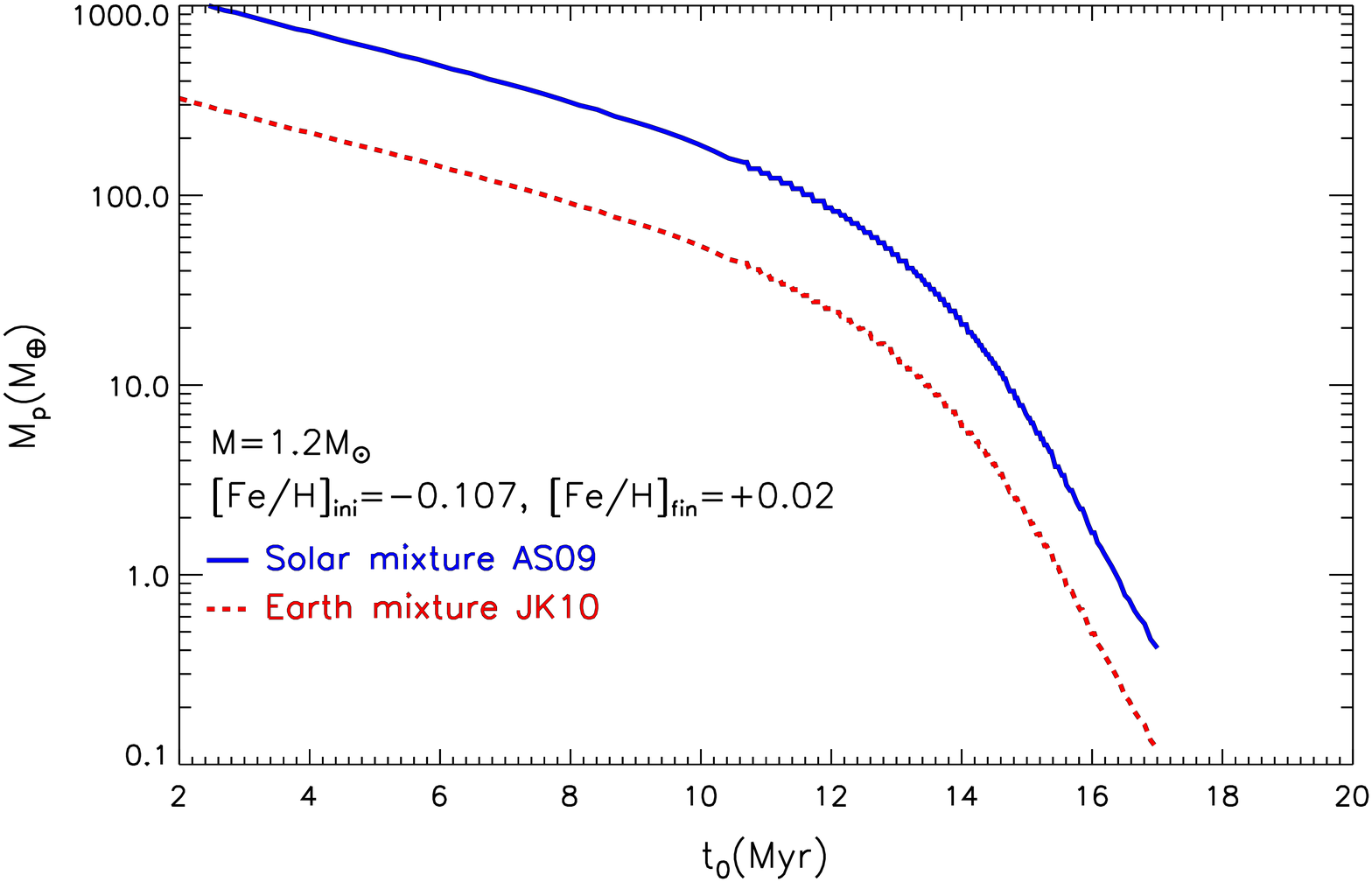}
\caption{Top panel: temporal evolution of the convective envelope mass extension for the standard model with $M=1.2$~\msun{} and [Fe/H]$=-0.107$. Bottom panel: ingested planet mass as a function of the age $t_0$ at which the ingestion occurs, for the Solar-like (AS09, solid blue line) and for the Earth-like heavy elements mixture (JK10, dashed red line).}
\label{fig:feh_mp}
\end{figure}
%% Figure: [Fe/H] fixed Mp
\begin{figure}
\centering
\includegraphics[width=\columnwidth]{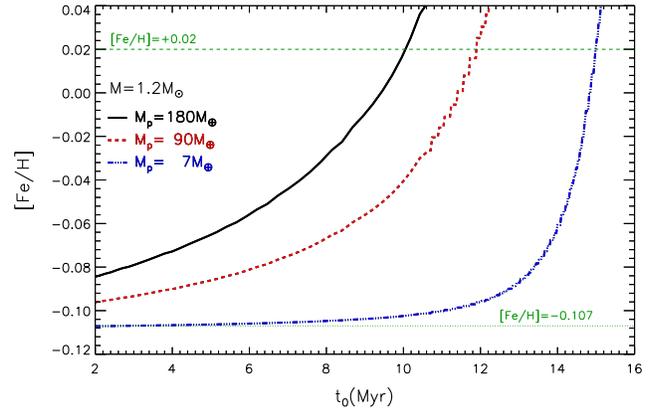}
\caption{Surface [Fe/H] value as a function of  $t_0$. Results for the three values of accreted mass $M_\rmn{p}=7$, 90 and 180~\mearth{} with the Solar-like AS09 metal distributions are shown.}
\label{fig:feh_mp_fix}
\end{figure} 

The mass of the ingested planet depends on the extension of the convective envelope of standard tracks, which, having fixed the stellar mass and initial chemical composition, is a function of the age of the star. Top panel of Fig.~\ref{fig:feh_mp} shows the temporal evolution the convective envelope mass extension during the \pms{} for our reference model (i.e. $M=1.2$~\msun{} and [Fe/H]$=-0.107$). In the selected time interval the convective envelope progressively reduces due to the radiative opacity decrease in the envelope. For this particular stellar mass and chemical composition, the surface convection vanishes at an age of about 17~Myr. 

Bottom panel of Fig.~\ref{fig:feh_mp} shows how the mass of the ingested planet changes, as a function of the age $t_0$ at which the accretion episode occurs. The computations have been performed for a star of 1.2~\msun{} which accretes matter with both the the Solar-like (AS09) and the Earth-like (JK10) heavy elements abundances. For a chosen planet chemical composition, the mass of the ingested planet necessary to account for the observed [Fe/H]$_\rmn{AS09}^{\#52}$ is proportional to the mass of the convective envelope $\Delta M_\rmn{c.e.}$ (see eq.~\ref{eq:mp}) which in turn varies with $t_0$. The earlier is the accretion (i.e. the smaller is $t_0$) and the thicker is the convective envelope and, consequently, the greater is the dilution of the accreted matter with the pristine stellar one and the larger is the requested planet mass $M_\rmn{p}$. Notice that for $t_0\ga 17$~Myr the convective envelope disappears and consequently the derived planet mass drops to zero.

For the selected stellar mass, we explored three values of $t_0$, namely 10, 12 and 15~Myr, the last approximatively corresponding to the cluster age. With this choice, in the case of the Solar-like mixture (AS09), we obtained from eq.~(\ref{eq:mp}) the following approximated values of the mass of the ingested planet, $M_\rmn{p}=180$ ($t_0=10$~Myr), 90 (12~Myr) and 7~\mearth{} (15~Myr). For the same $t_0$ values, $M_\rmn{p}$ reduces to $M_\rmn{p} = 53$ (10~Myr), 26 (12~Myr) and 2~\mearth{} (15~Myr), if the Earth-like mixture is adopted. The obtained  $M_\rmn{p}$ values are summarised in Table~\ref{tab:macc}.

We emphasize that the mass of the ingested planet -- at a give age $t_0$ and planet composition -- is fixed by the constraint [Fe/H]$^\rmn{new}$=[Fe/H]$^{\#52}$. Moreover, we also assumed that the initial mass of the accreting stellar model is fixed to 1.2~\msun. Thus, from eq.~(\ref{eq:mp}) the ingested planet mass $M_\rmn{p}$ depends only on 1) the age $t_0$ (through $\Delta M_\rmn{c.e.}$) and 2) on the Iron content of the accreted planet, i.e. $X_\rmn{Fe}^\rmn{p}$. 

If the chemical composition of the planet is fixed then the right side of eq.~(\ref{eq:mp}) is constant and independent of $t_0$. This implies that the ratio $M_\rmn{p}/\Delta M_\rmn{c.e.}$ has to be constant, too. A change of the age $t_0$ at which the accretion occurs directly translates into a change of the convective envelope extension $\Delta M_\rmn{c.e.}$, and the mass of the ingested planet $M_\rmn{p}$ must change accordingly, to keep the $M_\rmn{p}/\Delta M_\rmn{c.e.}$ ratio constant. In particular if $t_0$ increases, $\Delta M_\rmn{c.e.}$ decreases and $M_\rmn{p}$ must decrease too. This case corresponds to move along one of the rows of Table~\ref{tab:macc} (from left to right). 

On the other hand, at a fixed age $t_0$ -- fixed $\Delta M_\rmn{c.e.}$ -- the change of the planet chemical composition (i.e. $X_\rmn{Fe}^\rmn{p}$) modifies the right side of eq.~(\ref{eq:mp}) and the ratio $M_\rmn{p}/\Delta M_\rmn{c.e.}$ must change consequently. The use of the Earth-like mixture -- with an Iron content about 3.4 times larger than that in the Solar mixture -- leads to a reduction of the $M_\rmn{p}/\Delta M_\rmn{c.e.}$ ratio and, consequently, to a decrease of the ingested planet mass -- being $\Delta M_\rmn{c.e.}$ fixed. This case corresponds to move along one of the columns of Table~\ref{tab:macc}.
%%
%% Tab: Macc:
\begin{table}
\centering
\caption{Mass values of the ingested planet for the three selected ages $t_0$, obtained using eq.~(\ref{eq:mp}) and the Iron-poor and Iron-rich mixture.}
\label{tab:macc}
\begin{tabular}{rccc}
$t_0=$ & 10~Myr & 12~Myr & 15~Myr \\
\hline
\multicolumn{4}{c}{Iron-poor mixture: \citet{asplund09} Solar-like mixture}\\
$M_\rmn{p}$(M\earth) & 180 & 90 & 7 \\ 
\\
\multicolumn{4}{c}{Iron-rich mixture: \citet{javoy10} Earth-like mixture}\\
$M_\rmn{p}$(M\earth) & 53 & 26 & 2 \\ 
\hline
\end{tabular}
\end{table}

Fig.~\ref{fig:feh_mp_fix} shows how the final surface [Fe/H] changes as a function of the age $t_0$ at which the planet ingestion occurs in a star of 1.2~\msun{} and initial [Fe/H]=[Fe/H]$_\rmn{AS09}^\rmn{cl.}$. The figure shows the results obtained using eq.~(\ref{eq:feh}) with the three different values of the ingested planet, namely $M_\rmn{p} = 7$ ($t_0=15$~Myr), 90 (12~Myr) and 180~\mearth{} (10~Myr), adopting the AS09 Solar mixture for the planet. 

For a given $M_\rmn{p}$ there is a specific value $\overline{t}_0$ at which the planet must be ingested to produce the observed [Fe/H]$^{\#52}_\rmn{AS09}$. If the accretion begins earlier than $\overline{t}_0$, i.e. $t_0 \lt \overline{t}_0$, then the amount of metal-rich material supplied to the star is not enough to produce the requested surface [Fe/H]$^{\#52}_\rmn{AS09}$, as it is diluted in a thicker convective envelope. On the other hand, if the accretion starts later than $\overline{t}_0$, i.e. $t_0\gt \overline{t}_0$, then the resulting [Fe/H] is larger than the requested value as the accreted material is diluted in a thinner convective envelope. Consequently, having fixed the mass $M_\rmn{p}$, the final surface [Fe/H] strongly depends on the age $t_0$ at which the accretion occurs.

The results presented in Fig.~\ref{fig:feh_mp_fix} are unchanged if the Earth-like mixture (JK10) is adopted in place of the Solar-like one (AS09). The reason is that in both the cases, the mass of the ingested planet is chosen in order to contain the same total Iron content and to obtain exactly the same final [Fe/H].
 
%% Fig: [Fe/H], Mp vs mass, M.L.:
\begin{figure}
\centering
\includegraphics[width=\columnwidth]{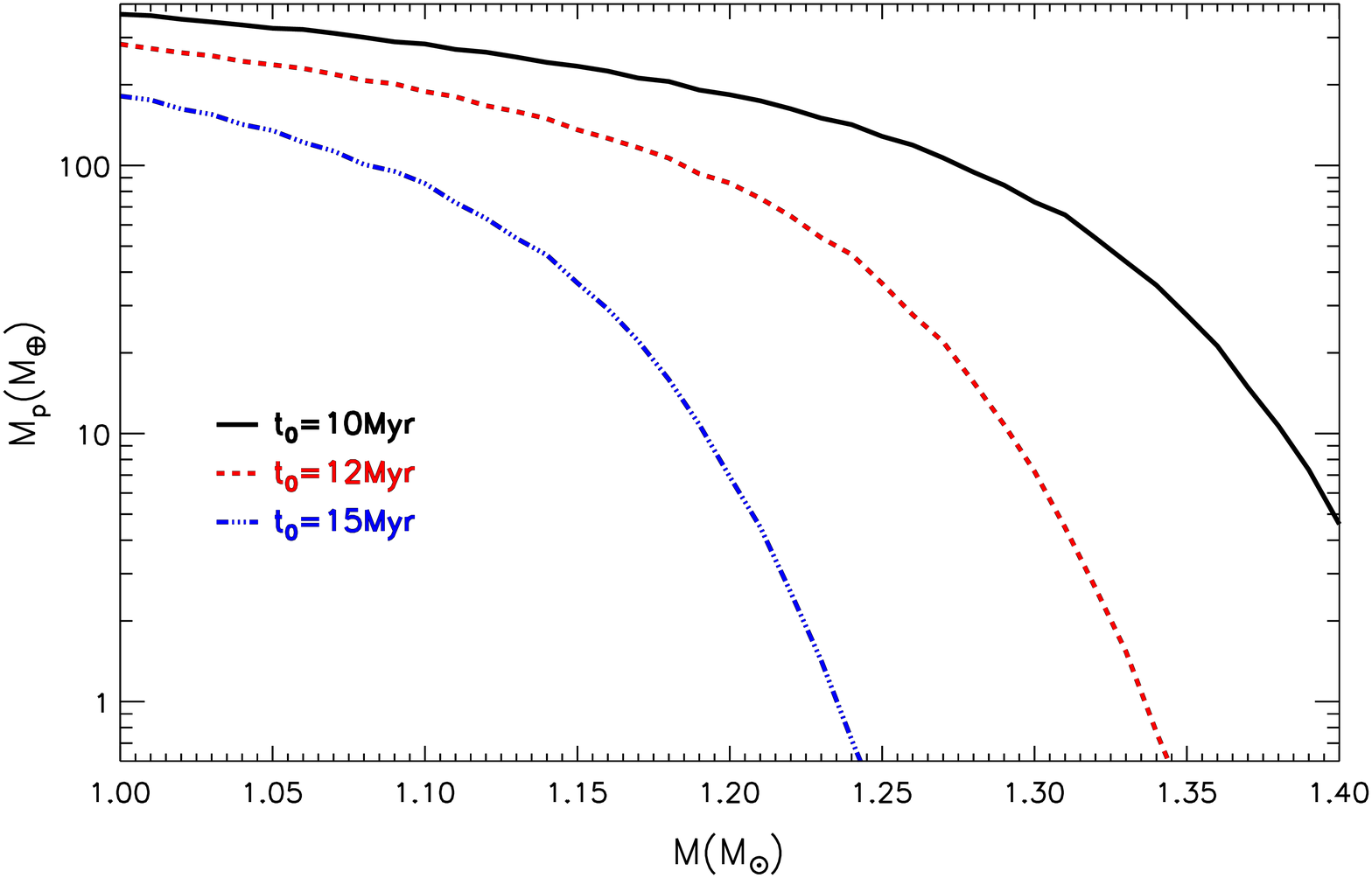}
\caption{Ingested planet mass $M_\rmn{p}$ as a function of the stellar mass for the simplified accretion model with [Fe/H]$=-0.107$, for $t_0=10,$ 12 and 15~Myr and the Solar-like (AS09) heavy elements mixture in the accreted matter.}
\label{fig:feh_mp_mas}
\end{figure} 

In the discussion above, we assumed that the stellar mass is fixed. This is a good assumption in our case because we focused on the analysis of a particular star ($\#52$) for which we have an estimate of both the age and the mass by fitting its \cmd{} location by means of our stellar tracks. However, it is worth to check the dependence of the presented results (i.e. mass of ingested planet) on the adopted stellar mass. 

Fig.~\ref{fig:feh_mp_mas} shows the derived ingested planet mass as a function of the stellar mass at a fixed $t_0$, in the mass range [1.0, 1.4]~\msun. As in the previous cases, we imposed to reproduce the surface [Fe/H]$^{\#52}_\rmn{AS09}$ starting from the initial [Fe/H] of the cluster. We performed the analysis for the same three values of the age at which the accretion occurs used in the previous discussion, namely $t_0=10$, 12 and 15~Myr.

As expected, at a fixed $t_0$, the value of $M_\rmn{p}$ changes with the stellar mass. We recall that the mass of the ingested planet derived in the simplified accretion model is proportional to the mass inside the convective envelope and, as well known, the larger is the stellar mass the smaller is the convective envelope at the same evolutionary stage during the \pms. Moreover, the evolutionary time scale of a star decreases by increasing its mass. Hence, for a fixed $t_0$, the convective envelope gets thinner and thinner (until it eventually disappears) as the mass increases, causing the rapid decrease of the derived $M_\rmn{p}$. Fig.~\ref{fig:feh_mp_mas} also shows that the decrease of $M_\rmn{p}$ vs. the stellar mass is amplified at increasing $t_0$. Thus, for a given $t_0$, the value of the derived ingested planet mass might significantly change by varying the mass of the star that is supposed to accrete.

If the Earth-like metal distribution is adopted in the ingested planet, the results are similar with the only difference that the mass of the planet is smaller by a factor $X_\rmn{Fe}(\rmn{JK10})/X_\rmn{Fe}(\rmn{AS09})\approx 3.4$.
 
\subsection{Full accretion models}
\label{sec:fullacc}
The simplified procedure described in the previous section and already adopted by \citet{spina14}, although straightforward to be implemented, is not fully consistent, as the accretion of metal-rich material necessarily modifies the stellar structure and evolution after the planet ingestion. The ingestion of metal-rich material and the consequent increase in the radiative opacity leads to a deeper convective envelope and a lower effective temperature. \citet{spina14} did not show the effect of properly treating the accretion process during the stellar evolution on the position of the track in the \cmd{} and,  consequently, on the inferred age and mass of the star. 

To analyse such an effect, we computed a second class of models by consistently following the accretion episode in stellar evolution calculation (hereafter \acc{} models). In this case, the star is evolved at constant mass with the initial [Fe/H]~=~[Fe/H]$^\rmn{cl.}_\rmn{AS09}=-0.107$ from the beginning of the Hayashi track to the age $t_0$, when the planet ingestion occurs.  Then, the mass is accreted, using exactly the same values of $t_0$, $M_\rmn{p}$ and chemical compositions quoted in Sect.~\ref{sec:simple}, with the accretion process consistently followed during the stellar evolution.

To do this we adopted the same accretion formalism described in \citet{tognelli15c}. The accretion is supposed to interest only a very small area of the star so that the most of the stellar surface is freely radiating the energy, allowing us to adopt the same boundary conditions (i.e. atmosphere models) used for the computation of non-accreting models. The mass is accreted only in the convective envelope where it is instantaneously mixed with the stellar matter contained in it. The accretion of matter modifies the whole stellar structure evolution because of the change of the total mass (gravitational energy) and of the chemical composition within the convective envelope. We discuss the results in the next Sections.

\section{Surface [Fe/H] and global metallicity evolution}
\label{sec:feh}
%%
%% Fig: [Fe/H], Mce -- acc:
\begin{figure}
\centering
\includegraphics[width=\columnwidth]{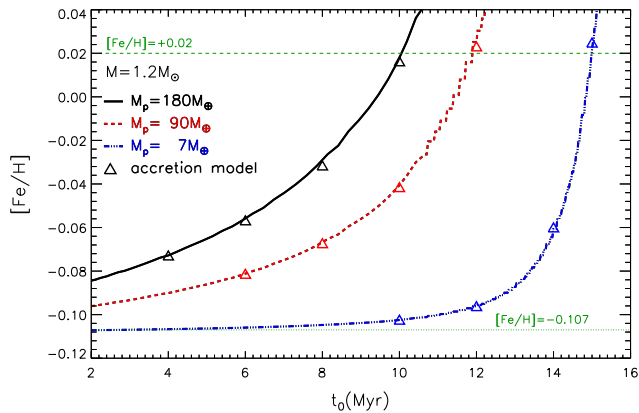}
\caption{The same as in Fig.~\ref{fig:feh_mp_fix} with over-plotted the [Fe/H] values obtained from the full accretion models after the accretion episode (empty triangles).}
\label{fig:feh_mp_acc}
\end{figure}
%%
%%
%%
%% Fig: [Fe/H], Mce -- acc:
\begin{figure}
\centering
\includegraphics[width=\columnwidth]{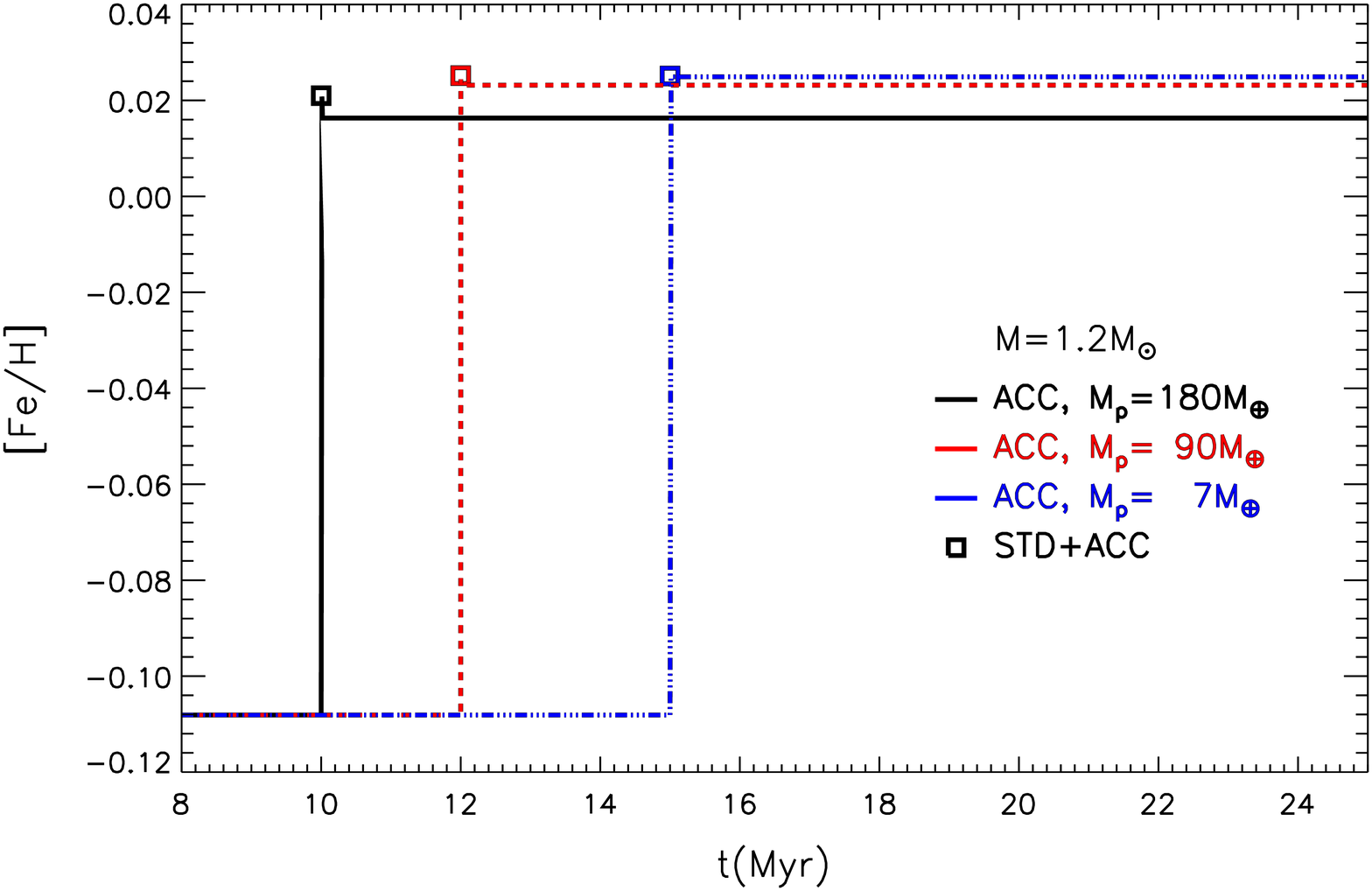}\\
\includegraphics[width=\columnwidth]{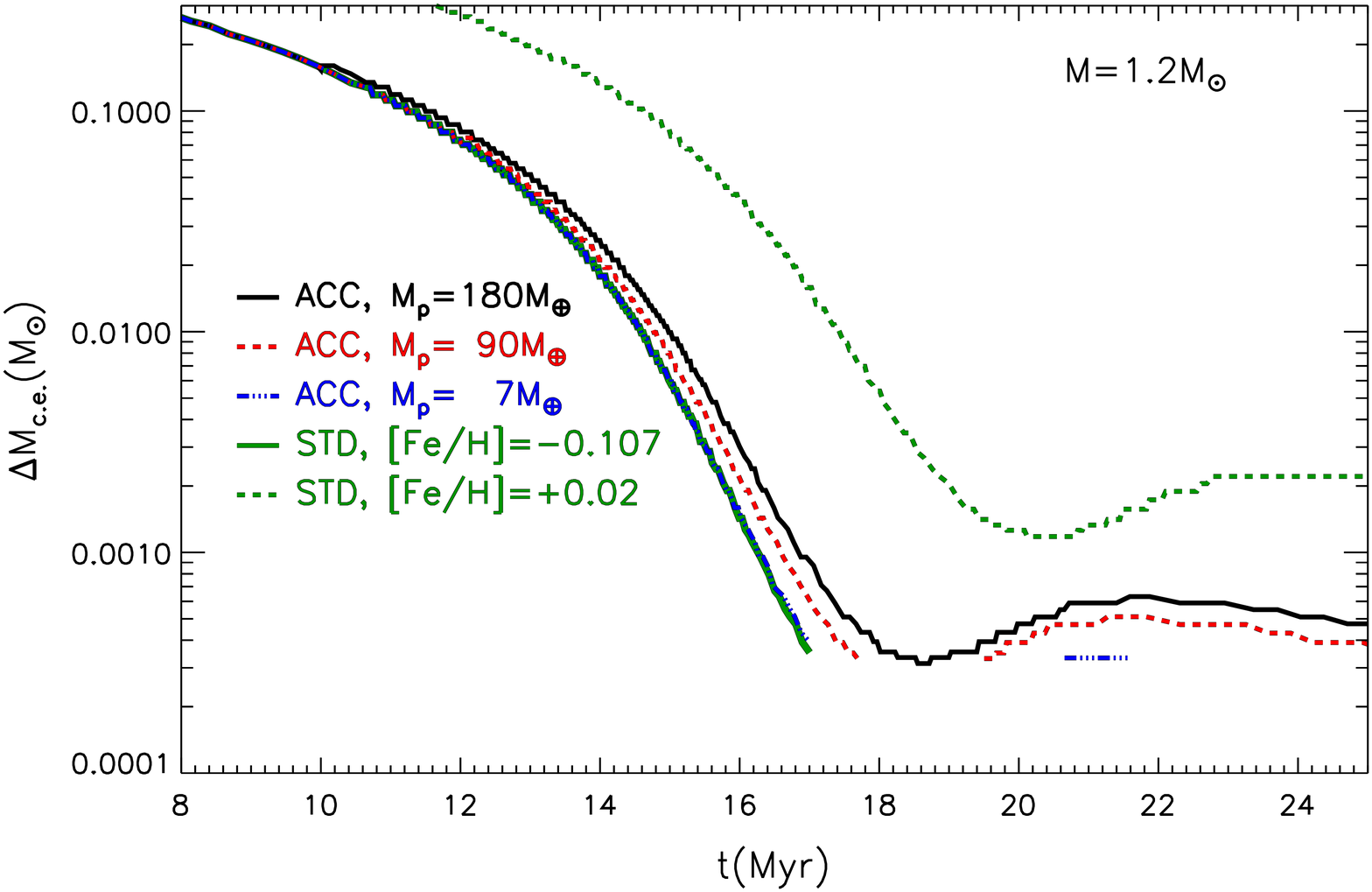}
\caption{Top panel: surface [Fe/H] value as a function of the stellar age for accreting models for the three values of the ingested planet mass ($M_\rmn{p}=7$, 90 and 180~\mearth) with the Solar-like AS09 metal distributions and three $t_0$ values, namely $t_0 = 10$ (180~\mearth), 12 (90~\mearth) and 15~Myr (7~\mearth). The empty squares correspond to the [Fe/H] obtained from the simplified accretion models discussed in Sect.~\ref{sec:simple} for the same $M_\rmn{p}$ and $t_0$. Bottom panel: temporal evolution of the mass inside the convective envelope ($\Delta M_\rmn{c.e.}$) for the same models shown in the top panel. Standard non accreting models with [Fe/H]=$-0.107$ and [Fe/H]$=+0.02$ are shown too (green lines).}
\label{fig:feh_mce_acc}
\end{figure}

In the simplified models described in Sect.~\ref{sec:simple} the surface Iron abundance evolution of a planet-ingesting star is a simple step function, with [Fe/H]=[Fe/H]$_\rmn{low}$ for $t\lt t_0$ and [Fe/H]=[Fe/H]$_\rmn{high}$ for $t\ge t_0$. However, as mentioned in Sect.~\ref{sec:fullacc}, the surface Iron abundance evolution occurring in fully accretion models computation is more complex as the planet ingestion modifies also the stellar structure evolution. 

Fig.~\ref{fig:feh_mp_acc} shows the surface [Fe/H] after the accretion episode obtained using the full accretion model, computed for different values of $t_0$. We performed the analysis using the three $M_\rmn{p}$ used in the simplified accretion case, namely 7, 90 and 180~\mearth{}(AS09 solar mixture). The results of the full accretion model is compared to the predictions of the simplified one with the same $M_\rmn{p}$. As already discussed, for a given $M_\rmn{p}$ there is only one specific value of the accretion age $\overline{t}_0$ that produces the requested final [Fe/H]=[Fe/H]$^\rmn{\#52}$. We computed the surface [Fe/H] for full accreting models at three $t_0 \lt \overline{t}_0$ and at $t_0=\overline{t}_0$. Figure shows that the surface [Fe/H] after the planet ingestion obtained using the simplified accretion model fully agrees with that predicted by the detailed accretion one. Slight differences are present only in the case of large planets accretion, $M_\rmn{p} \ga 90$~\mearth. In this case the detailed calculations show a tendency of producing a lower [Fe/H] after the accretion episode, although the difference is always smaller than 0.01~dex. 

To better analyse such a difference, we focused on the planet ingestion at $t_0=\overline{t}_0$, i.e. the age needed to produce the requested [Fe/H]=[Fe/H]$^{\#52}$. Top panel of Fig.~\ref{fig:feh_mce_acc} shows the surface [Fe/H] for models that fully account for the accretion process (\acc{} models). We adopted exactly the same values of the accretion age $t_0$ and of the mass of the ingested planet $M_\rmn{p}$ used in the simplified models for the solar heavy element mixture (AS09), namely $M_\rmn{p}/$\mearth =180 (10~Myr), 90 (12~Myr) and 7 (15~Myr). For comparison, in the figure we over-plotted the [Fe/H] values obtained by the simplified accretion model at the same age $t_0$ and for the same $M_\rmn{p}$. 

The surface Iron abundance evolution (top panel) is the same as the simplified models for $t\lt t_0$, since before the planet ingestion the star evolves as a standard \pms{} model without accretion, and consequently [Fe/H]=[Fe/H]$_\rmn{low}$. On the other hand, for $t=t_0$, when the planet ingestion occurs, [Fe/H]=[Fe/H]$_\rmn{high}$. For $t\gt t_0$ the accreting models show a quick -- even if modest -- decrease of the surface [Fe/H], which is more pronounced for large $M_\rmn{p}$. Such a behaviour is a peculiarity of the models that actually follow the accretion process. It is the result of the quick increase of the opacity in the external convective region -- caused by the accretion of metal-rich material -- which makes the bottom of the convective envelope to sink inwards, thus reaching regions with the pristine chemical composition. Consequently, the accreted metal-rich material is further diluted leading to the observed modest  decrease in the surface [Fe/H], with differences that grow by increasing the ingested planet mass. However, the maximum extent of such differences is small, being less than 0.01~dex, and consequently negligible if compared to the uncertainty on abundance determinations. To be noted that, after the accretion and after the quick drop of [Fe/H], the surface abundances remain constant because the convective envelope never extends deep enough to reach regions not interested by the accretion and no further dilution occurs. 

Bottom panel of Fig.~\ref{fig:feh_mce_acc} shows the temporal evolution of the mass contained inside the convective envelope for the same accreting (\acc) models shown in the top panel. For comparison, we also plotted the standard (\std) model with the low- and high- initial [Fe/H], i.e [Fe/H]$^\rmn{cl.}_\rmn{AS09}=-0.107$  and [Fe/H]$^{\#52}_\rmn{AS09}=+0.02$. As well known, the more metallic is the structure the thicker is the external convective envelope at the same evolutionary stage. 

The differences between the accreting and standard sets of models get larger and larger by increasing the mass of the ingested planet. The model that actually follows the accretion process with the lowest value of $M_\rmn{p}$ (i.e. $M_\rmn{p}=7$~\mearth) shows an extension of the convective envelope almost indistinguishable from that of the low-[Fe/H] standard one. On the other hand, by increasing $M_\rmn{p}$ the convective envelope gets -- after the planet ingestion -- progressively thicker and thicker with respect to the standard low-[Fe/H] model. 
Just after the accretion the mass extension of the convective envelope can change of about 10 percent if a 180~\mearth{} planet is accreted and of about 6--4 percent for a 90 and 7~\mearth{} planet. The differences between the extension of the convective envelope in standard and accretion models increase at larger ages. 

The different evolution of the convective envelope in presence of accretion can be explained by considering two aspects. Firstly, the accretion of the metal-rich material with the AS09 mixture produces a relevant increase of the Iron content (i.e. the [Fe/H] value) but also of the total metallicity $Z^\rmn{new}$ in the envelope (see e.g. eq.~\ref{eq:zsup}) to about $Z^\rmn{new} \approx 0.0135$--0.0138 (i.e. increased by about 30--35~percent). These values have to be compared with those of both the low-[Fe/H] ($Z^\rmn{ini}=0.0102$) and the high-[Fe/H] (i.e. $Z^\rmn{ini}=0.0135$) models. It is evident that the resulting surface $Z^\rmn{new}$ of the AS09 models is essentially the same of the high-[Fe/H] metallicity standard ones. Secondly, if the accretion begins earlier (smaller $t_0$ and larger $M_\rmn{p}$) the star has a thicker convective envelope and the modification of the chemical composition due to the accretion of metal-rich material occurs on a larger fraction of the star. Consequently, as $t_0$ decreases, the accreting model gets progressively more similar to the standard one with the high-[Fe/H] value in the whole structure. On the contrary, if the accretion begins later (larger $t_0$ and smaller $M_\rmn{p}$) when the convective envelope is thinner, a larger part of the star has a chemical composition equal to that of the standard models with low-[Fe/H], while the enhanced metals abundance is confined to the outer region. As such, the accreting model is expected to be much closer to the standard low-[Fe/H] one.

We want to emphasize that the final metallicity $Z^\rmn{new}$ in the convective envelope -- after the planet ingestion -- does not depend neither on $t_0$ nor on $M_\rmn{p}$, once the chemical composition of the planet has been fixed, as a consequence of the constraint [Fe/H]$^\rmn{new}$=[Fe/H]$^{\#52}$ imposed to compute $M_\rmn{p}$. This can be easily seen using eq.~(\ref{eq:zsup}). Noticing that $M_\rmn{p}/\Delta M_\rmn{c.e.} \approx 10^{-3}$--$10^{-4}$ at $t_0$, one obtains 
\begin{equation}
Z^\rmn{new}\approx Z  + Z^\rmn{p}\frac{M_\rmn{p}}{\Delta M_\rmn{c.e.}}
\label{eq:zz}
\end{equation}
As discussed in Sect.~\ref{sec:simple}, having fixed $X_\rmn{Fe}^\rmn{p}$ the ratio $M_\rmn{p}/\Delta M_\rmn{c.e.}$ is fixed and $Z^\rmn{new}$ is independent of $M_\rmn{p}$ and $t_0$. 

Notice that, close to the \zams{} -- i.e. for $t\ga 20$~Myr -- the convective envelope of accreting models with $M_\rmn{p} \ga 90$~\mearth{} shows an oscillation similar to that of standard models with the high-[Fe/H], which leads to a slight increases of $\Delta M_\rmn{c.e.}$. However, this temporarily oscillation does not affect the surface chemical composition, as the maximum depth reached by the surface convection during the oscillation is much smaller than that reached by the surface convection at the moment of the accretion. Consequently, the convective mixing is occurring in a region of homogeneous chemical composition, already mixed during the previous evolution after the accretion.

In order to study the impact of the chemical composition of the ingested planet on the final stellar surface elements abundance we performed the same analysis discussed above but using the Earth-like mixture (JK10) for the accreted material. In this case, the mass abundance of Iron $X_\rmn{Fe}^\rmn{p}$ is about 3.4 times larger than that in the Solar-like one. However, we recall that for a given accretion age $t_0$, the total mass of the ingested planet is fixed by the constraint that the final [Fe/H] must match the observed [Fe/H]$^{\#52}_\rmn{AS09}$. This, as shown in Sect.~\ref{sec:simple}, results in an ingested planet mass $M_\rmn{p}$ smaller than that obtained for the Solar-like mixture. 

The surface [Fe/H] evolution in the case of Solar-like or Earth-like heavy element mixture is similar, as the total mass of accreted Iron is forced to be the same. On the other hand, this is not true for the final metallicity $Z^\rmn{new}$ in the convective envelope. The smaller planet mass $M_\rmn{p}$ required in the case of the Earth-like mixture with respect to that of the Solar-like one at the same age $t_0$ -- i.e. at the same $\Delta M_\rmn{c.e.}$ -- causes a smaller increase of the surface metallicity $Z^\rmn{new}$. This can be easily understood by looking at eq.~(\ref{eq:zz}), as the metallicity variation is proportional to the mass of the ingested planet $M_\rmn{p}$. The use of the Earth-like mixture (JK10) in the accreted matter leads to models with a maximum total metallicity in the envelope of about $Z^\rmn{new}=0.0112$ (i.e. increased by about 10~percent), a value which is very close to that used for the computation of the standard low-[Fe/H] models, namely $Z^\rmn{ini}=0.0102$, and much lower than that of the high-[Fe/H] ones ($Z^\rmn{ini}=0.0135$). Thus, even if the Iron content in the envelope is significantly increased ([Fe/H]$\approx +0.02$) the total metallicity $Z^\rmn{new}$ is only marginally affected by the accretion. As a consequence, such models show an evolution of $\Delta M_\rmn{c.e.}$ very similar to that of the standard low-[Fe/H] ones, with a maximum variation of $\Delta M_\rmn{c.e.}$ of about 4 percent when a 53~\mearth{} planet is ingested.

From this discussion it emerges a first important result. The adoption of an Iron-poor or Iron-rich mixture for the accreted matter has the effect of producing two limiting sets of models: one with an high-metallicity $Z^\rmn{new}$ similar to that of the high-[Fe/H] standard models (Iron-poor AS09 mixture), and the other with a low-metallicity more similar to that of the low-[Fe/H] ones (Iron-rich JK10 mixture, rocky planet). 

\section{Surface heavy element abundances}
\label{sec:heavy}
%% Figure: chemical abundances:
\begin{figure}
	\centering
	\includegraphics[width=\columnwidth]{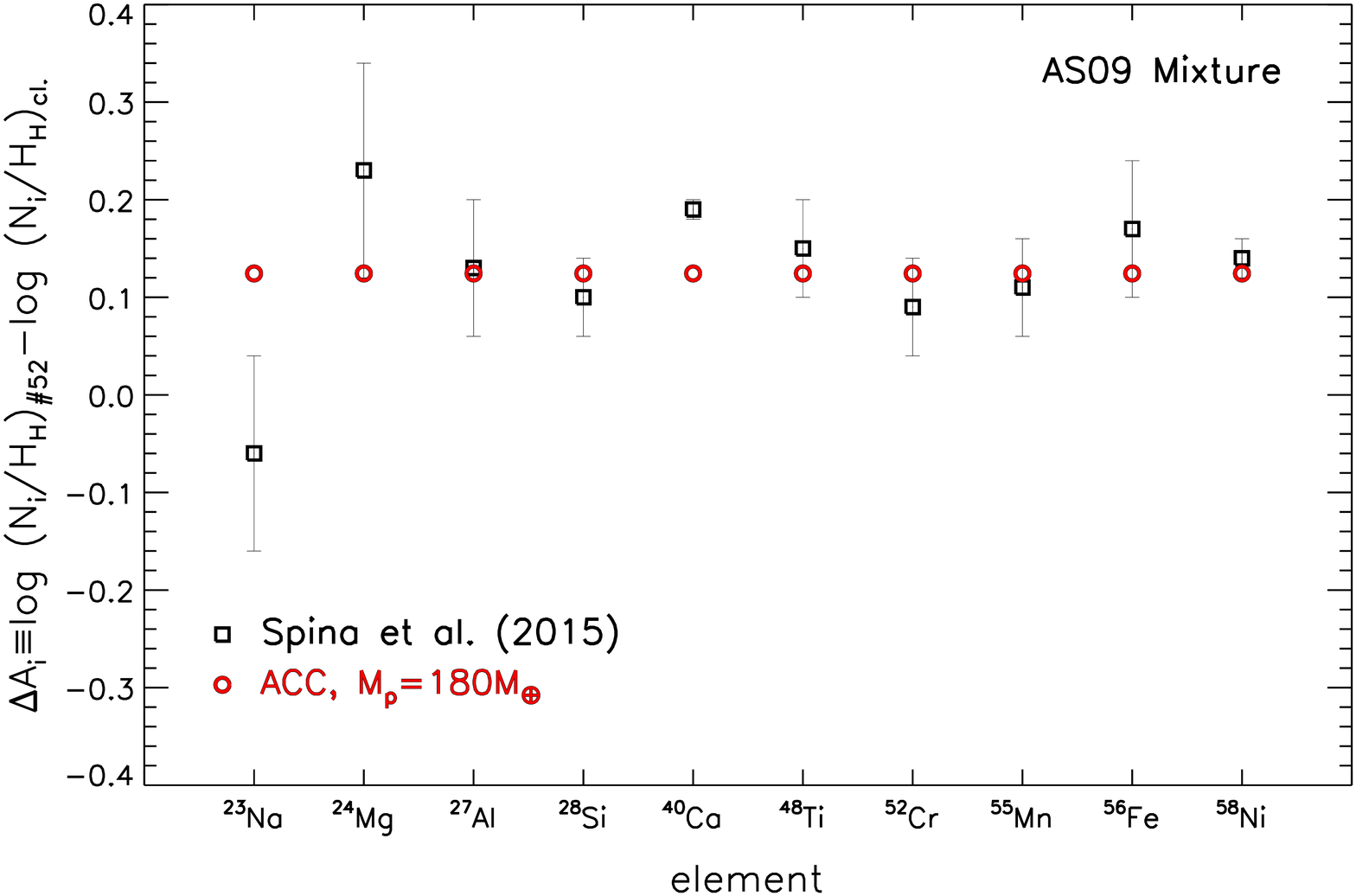}
	\includegraphics[width=\columnwidth]{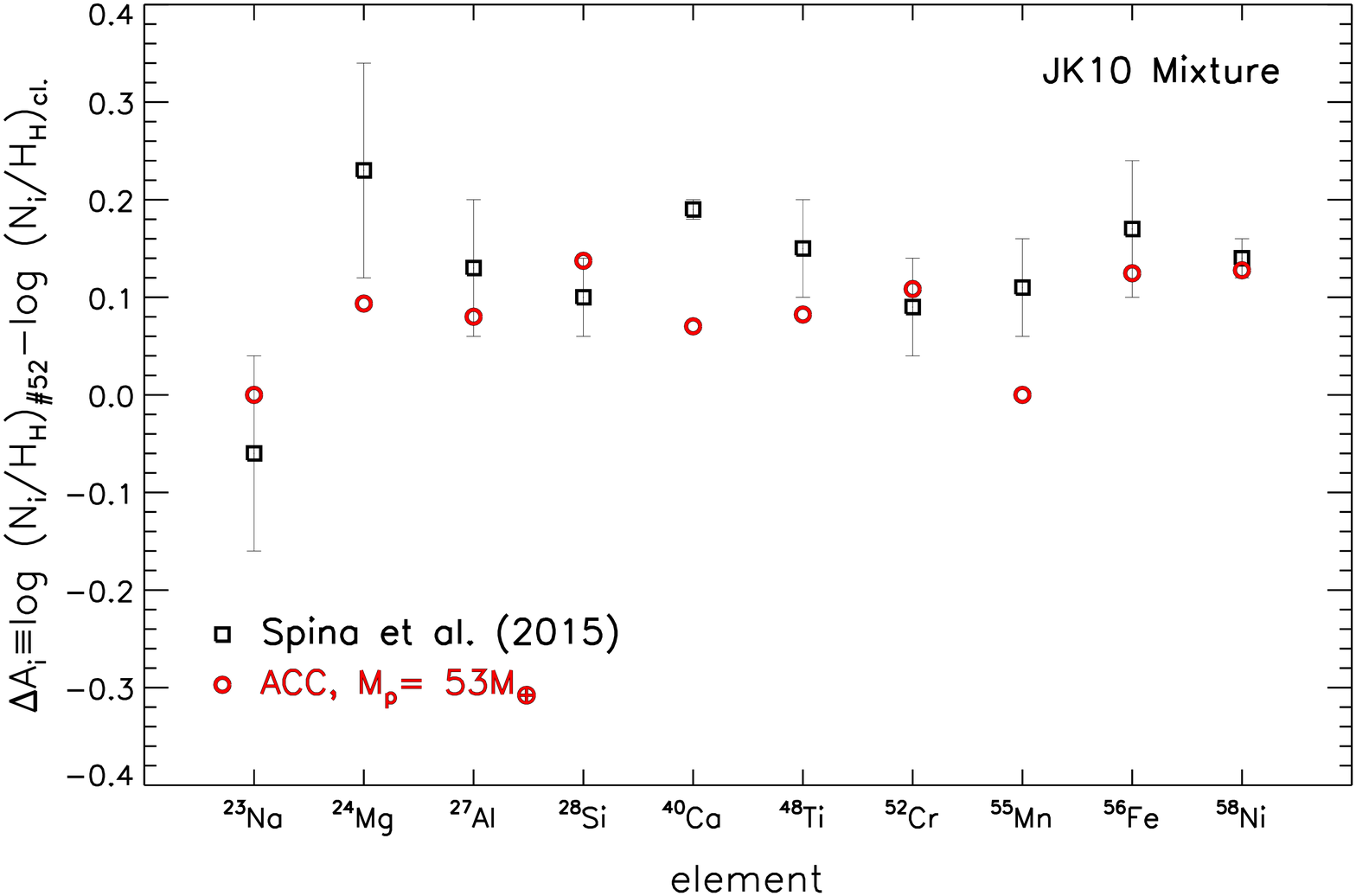}
	\caption{Observed (black squares) and predicted (red circles) surface abundance differences $\Delta A_\rmn{i}$  between the peculiar star $\#52$ and the mean cluster values for the labelled elements. Top panel: predictions of full accretion models obtained using the Solar-like AS09 mixture in the accreted matter. Bottom panel: the same as top panel but for models with the Earth-like JK10 mixture in the accreted matter.}
\label{fig:chm_ele}
\end{figure}
Another point to discuss is the predicted surface elements abundances after the accretion episode. To this regard, \citet{spina15} measured the abundances of several elements (Na, Mg, Al, Si, Ca, Sc, Ti, V, Cr, Mn, Fe, Co, Ni Cu) in both the peculiar star $\#52$ and in another cluster star, namely $\#45$, representative of the mean cluster composition. They provided the differences of the surface abundances between these two stars, i.e.
\begin{equation}
 \Delta A_\rmn{i}\equiv \log (N_\rmn{i}/N_\rmn{H})_{\#52} - \log (N_\rmn{i}/N_\rmn{H})_{\#45}
\end{equation}
The variation of the surface elements abundance depends on the characteristics of the accretion episode, in particular on the chemical composition of the accreted matter. Thus, it is worth to compare the predicted $\Delta A_\rmn{i}$ obtained in our computations with those measured by \citet{spina15}. 

Fig.~\ref{fig:chm_ele} shows the comparison between the predicted and observed $\Delta A_\rmn{i}$ for a sub-sample of elements available both in our models and in \citet{spina15}. The upper panel shows the case for the Solar-like mixture and the bottom panel the Earth-like one. Notice that, the final surface abundances are independent of $M_\rmn{p}$ if the ingested planet mass is tuned to obtain the same [Fe/H] after the accretion episode. 

In the case of the Solar-like mixture, the results of the accretion models agrees quite well with the values measured by \citet{spina14}, within the observational uncertainties, with the exception of Na (over-estimated in our models) and Ca (under-estimated in our models). Nevertheless, the adoption of a Solar-like mixture seems to be a good approximation, at least within the current uncertainties, of the metals distribution in the accreted matter. Since the heavy element mixture used in the star and in the accreted matter is the same, the accretion produces an abundance enhancement of all the elements by the same factor. 

The situation is different if the Earth-like mixture is used for the planet. In this case, the enhancement factor is different for each element. If the Earth-like mixture is adopted, the agreement with the observed and predicted $\Delta A_\rmn{i}$ pattern gets slightly worst, in particular for Mg, Ti and Mn. For all these elements we under-estimated the abundance, although in almost all the cases we are still compatible within 2 sigma. 

We performed the same analysis using the simplified accretion scheme that, as expected, produces abundance very close (with differences lower than 0.01~dex) to the full accretion one.

These comparisons show that the accretion models are compatible with the observed $\Delta A_\rmn{i}$ pattern using both the Solar-like and Earth-like mixture in the accreted matter. More accurate surface abundances are needed to better constrain the chemical composition of the ingested planet. We remark that our aim is not to reproduce exactly the observed abundances but rather to investigate the effect of adopting different values of the not well constrained parameters, such as the planet mass and its chemical composition, that significantly affect the stellar models after the planet ingestion.

\section{Effect on the Colour-Magnitude Diagram}
\label{sec:cmd}
%%
%%
%% Fig: CMD -- acc:
\begin{figure}
\centering
\includegraphics[width=\columnwidth]{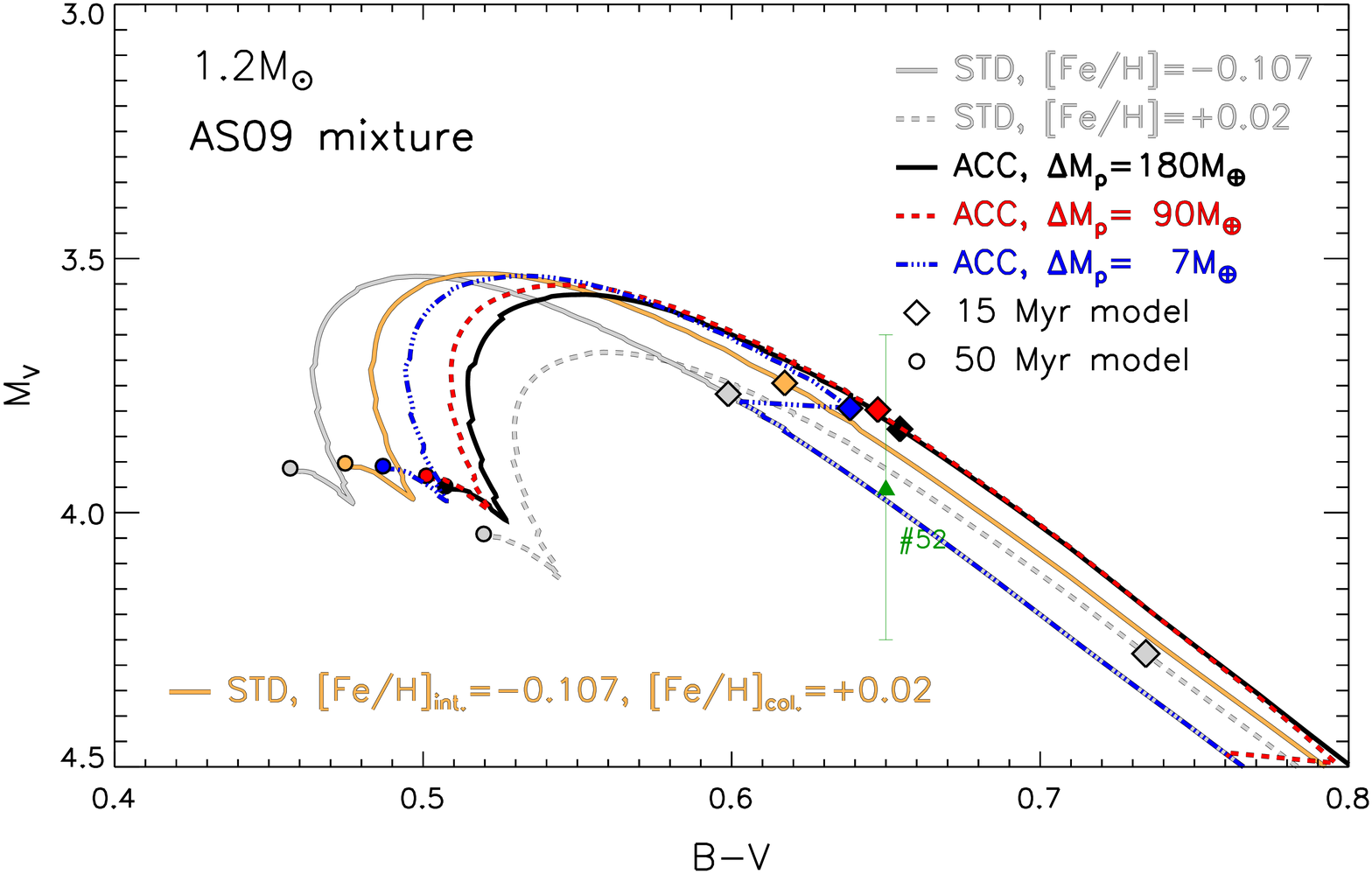}\\
\includegraphics[width=\columnwidth]{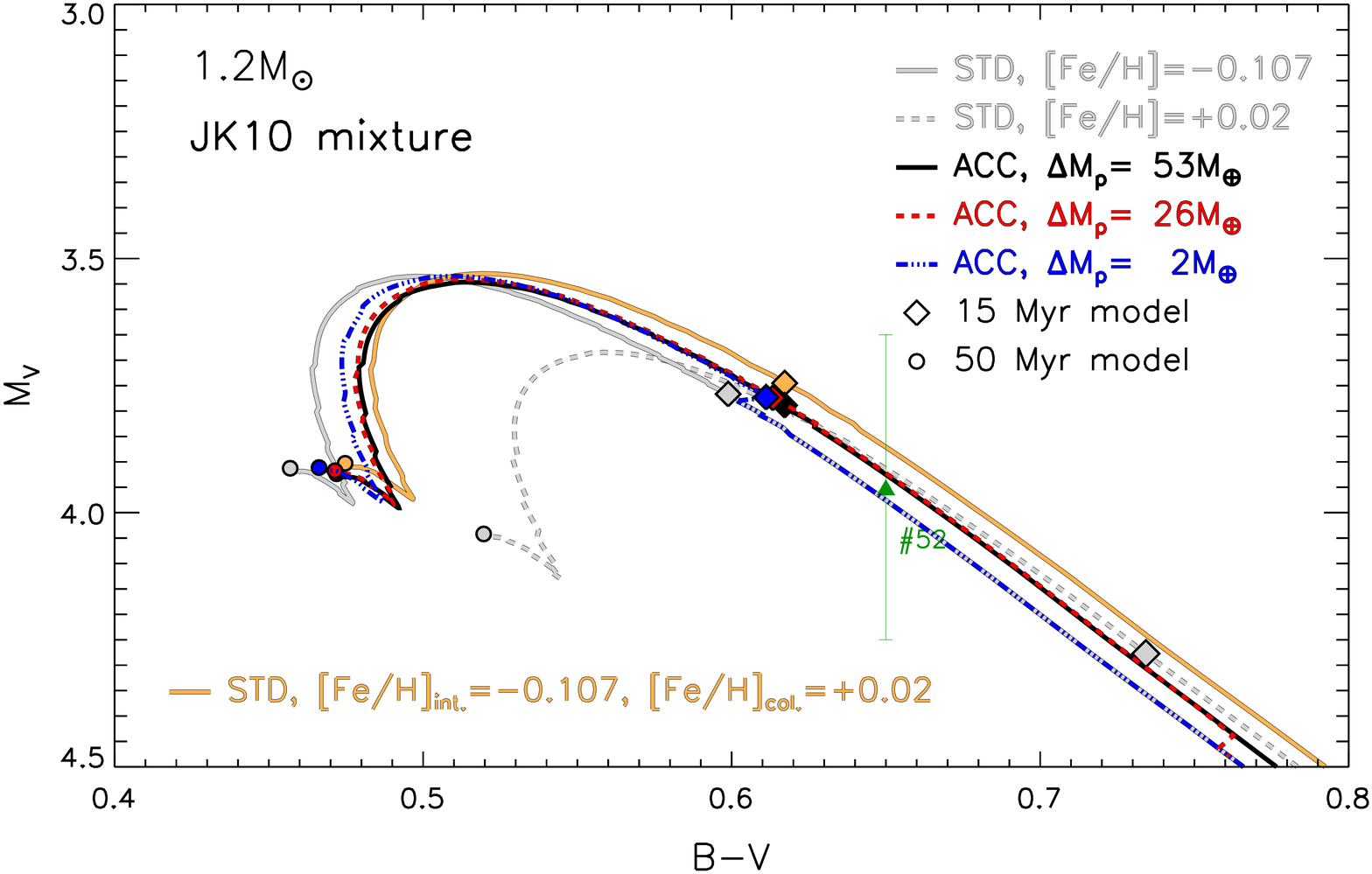}
\caption{Colour-Magnitude diagram with the standard low-[Fe/H] (\std, [Fe/H]$=-0.107$, grey solid line), standard high-[Fe/H] (\std,  [Fe/H]$=+0.02$, grey dashed line), and accreting evolutionary tracks (\acc, coloured lines). The \acc{} tracks are computed for the three labelled $M_\rmn{p}$. The standard evolutionary track computed for [Fe/H]=$-0.107$ and converted into the CMD using a surface [Fe/H]=[Fe/H]$^{\#52}=+0.02$ is also shown. The models corresponding to an age of 15~Myr (filled diamonds) and 50~Myr (filled circles) are shown too, together with the position of $\#52$. Top panel: models computed assuming the \citet[][AS09]{asplund09} Solar-like heavy elements distribution in the accreted matter. Bottom panel: the same as in top panel, but for the \citet[][JK10]{javoy10} Earth-like metals distribution.}
\label{fig:cmd_acc}
\end{figure}
%%
%%
%% Tab: model properties AS09/JK10:
\begin{table}
\centering
\caption{Main characteristics of the 15~Myr and 50~Myr standard and accreting models shown in Fig.~\ref{fig:cmd_acc}.}
\label{tab:tab_cmd}
\begin{tabular}{lccccc}
\multicolumn{6}{c}{age=15 Myr}\\
\hline
type & $M_\rmn{p}$& $B-V$ & $M_V$ & [Fe/H] & mixture\\
     &  (\mearth)\\
\hline
\hline
\std &   0 & 0.5989 & 3.77 & $-0.107$ & --\\
\std &   0 & 0.7343 & 4.28 & $+0.020$ & -- \\
\acc$^{a}$ &   7 & 0.6384 & 3.79 & $+0.025$ & AS09\\
\acc$^{a}$ &  90 & 0.6474 & 3.80 & $+0.023$ & AS09\\
\acc$^{a}$ & 180 & 0.6546 & 3.84 & $+0.016$ & AS09\\
\acc$^{b}$ &   2 & 0.6112 & 3.77 & $+0.025$ & JK10\\
\acc$^{b}$ &  26 & 0.6134 & 3.77 & $+0.024$ & JK10\\
\acc$^{b}$ &  53 & 0.6172 & 3.79 & $+0.017$ & JK10\\
\std+\acc$^{c}$ &   0 & 0.6172 & 3.74 & $+0.020$ & --\\
\hline
\\
\multicolumn{6}{c}{age=50 Myr}\\
\hline
type & $M_\rmn{p}$& $B-V$ & $M_V$ & [Fe/H] & mixture\\
     &  (\mearth)\\
\hline
\hline
\std &   0 & 0.4569 & 3.91 & $-0.107$ & --\\
\std &   0 & 0.5196 & 4.04 & $+0.020$ & --\\
\acc$^{a}$ &   7 & 0.4870 & 3.91 & $+0.025$ & AS09\\
\acc$^{a}$ &  90 & 0.5011 & 3.93 & $+0.023$ & AS09\\
\acc$^{a}$ & 180 & 0.5072 & 3.95 & $+0.016$ & AS09\\
\acc$^{b}$ &   2 & 0.4663 & 3.91 & $+0.025$ & JK10\\
\acc$^{b}$ &  26 & 0.4714 & 3.92 & $+0.024$ & JK10\\
\acc$^{b}$ &  53 & 0.4720 & 3.92 & $+0.017$ & JK10\\
\std+\acc$^{c}$ &   0 &  0.4747 & 3.90 & $+0.020$ & --\\
\hline
\end{tabular}
\medskip
\flushleft
$^{a}$ accretion of matter with the AS09 mixture.\\
$^{b}$ accretion of matter with the JK10 mixture.\\
$^{c}$ standard low-[Fe/H] track with high-[Fe/H] colors.\\
\end{table}
We showed that the ingestion of a planet affects not only the surface chemical abundance of a star but also the structure evolution. Thus, the planet engulfment is expected to produce effects also on the track morphology/position in the \cmd, mainly because of 1) the variation of the luminosity and effective temperature caused by the accretion process and 2) the metallicity change in the atmosphere. The first effect can be obtained only by means of the full accretion model, while the second can be evaluated using the simplified approach. 

Fig.~\ref{fig:cmd_acc} shows the evolutionary tracks of $M=1.2$~\msun{} in the ($B-V$, $M_V$) plane. The standard models with the two extreme [Fe/H] values, namely [Fe/H]$^\rmn{cl.}_\rmn{AS09}=-0.107$ and [Fe/H]$^{\#52}_\rmn{AS09}=+0.02$, and the accreting models computed adopting the three quoted planet masses $M_\rmn{p}$ for the Solar-like (AS09, top panel) and the Earth-like mixtures (JK10, bottom panel) in the accreted matter are shown. We over-plotted a standard track computed with the mean cluster [Fe/H] (in the interior), but converted into the colour-magnitude plane using a the $\#52$ surface [Fe/H]$=+0.02$. This model corresponds to take into account only the effect of varying the chemical composition in the stellar atmosphere (i.e. simplified accretion model). For all the evolutionary sequences, we also marked the position in the CMD of the models at the age $t=15~$Myr (approximatively the age of the  Gamma Velorum cluster) and $t=50~$Myr (on the \ms). The main characteristics of the 15 and 50~Myr models are listed in Table~\ref{tab:tab_cmd}. In figure we also over-plotted the observational data for $\#52$ corrected for the reddening and distance modulus given in \citet{spina14}. Notice that at the age of the cluster (15~Myr) the location of the standard model computed adopting the initial low-[Fe/H]$=-0.107$ is about 0.05~mag bluer (in colour) and 0.1--0.2 mag brighter than the data. However, if an age of 14~Myr is adopted, the low-[Fe/H] model matches the observed position of $\#52$ quite well\footnote{If the 1.2~\msun{} model with the high-[Fe/H]=$+0.02$ is adopted, the best age is slightly larger, about 16.5~Myr, but still compatible with the cluster age.}.

The standard track with the initial high-[Fe/H] is the coolest and fainter while that with the initial low-[Fe/H] is the hottest and brighter. The non-standard ones computed taking into account the accretion episode are located in between. The inclusion of the accretion affects the position of the track in the \cmd, depending on the evolutionary phase, mass $M_\rmn{p}$ and the chemical composition of the ingested planet. As already stated, the main effect of the accretion is to increase the [Fe/H] and the metallicity inside the convective envelope leaving unchanged the metallicity in the stellar core. Such an increase of the metal content causes a shift of the track towards lower effective temperatures altering only marginally the total luminosity and the temporal evolution, which are more sensitive to the core metallicity. In addition, the accreting models get redder and redder as $M_\rmn{p}$ increases (see Table~\ref{tab:tab_cmd}), as a consequence of the progressively larger $Z$ in a larger fraction of the star. However, such a metallicity variation produces an effect that might be not clearly visible in all the evolutionary phases. As an example, the three accreting models are almost coincident during the Henyey phase of the \pms{} evolution\footnote{The Henyey track corresponds to that part of the \pms{} evolution from the formation of a sizeable radiative core to the maximum in luminosity before the \zams, when the star moves towards higher effective temperatures.}. During this phase the shift induced by the accretion is almost parallel to the evolutionary track. On the other hand, the separation gets more and more visible as the star approaches the \zams. This fact clearly shows that the effect of a planet ingestion on the structure of the star is not limited to the period just after the accretion itself, but it is still present  much later, i.e. near the \ms. In other words, the planet ingestion leaves footprints that can be observed even long after the accretion. We only mention that, for ages larger than about 50--100~Myr (i.e. on the MS), the metals over-abundance in the convective envelope can be erased by non-standard turbulent mixing \citep[thermohaline mixing, see e.g.][]{vauclair04} and the surface chemical anomalies might significantly reduce \citep[see e.g.][]{garaud11,theado12}.

The largest differences between the standard low-[Fe/H] and the accreting models occur when it is ingested a planet of 180~\mearth{} with the Solar-like heavy elements mixture (AS09, top panel), because of the relatively large increase of the surface metallicity $Z$ (about $30$~percent) with respect to the initial one. Referring to the 15~Myr model location in the CMD, we obtained a significant reduction of the effective temperature of the accreting model of about 120~K, which corresponds to a track redder of about 0.05~mag than the standard low-[Fe/H] one. Similar variation are obtained for the 50~Myr model (see Table \ref{tab:tab_cmd}). Notice that at the age of 15~Myr all the three accretion models match quite well the observed position of $\#52$ in the \cmd.

In the case of an Earth-like planet ingestion (JK10, bottom panel), the maximum variation of the surface metallicity $Z$ is much smaller (about $10$~percent) and the accreting tracks are very close to the low-[Fe/H] standard ones, even when the higher value of the ingested planet mass is considered (i.e. 53~\mearth). We obtained an accreting track cooler than about 40~K and redder of about 0.01--0.02~mag than the low-[Fe/H] standard one for both the age of 15 and 50~Myr (see Table \ref{tab:tab_cmd}). In this case the 15~Myr accretion models are slightly hotter and brighter than $\#52$ in the \cmd. However, such models match the data if an age of 14~Myr is adopted.

We also analysed the effect on the CMD position of a standard track with the initial [Fe/H] of the cluster, converted into magnitude and colors using a surface [Fe/H] = [Fe/H]$^{\#52}=+0.02$. This case corresponds to the simplified accretion scheme, where we assumed that the accretion modifies only the surface chemical composition of the star in a very thin layer so that the structure (mainly $T_\rmn{eff}$ and luminosity) is unaltered. The adoption of an higher [Fe/H] in the atmospheric models leads to an almost rigid translation of the track in the CMD, with a shift of $+0.02$~mag in colour and less than $-0.01$~mag in magnitude, without changing the temporal evolution (because no structural change occurs). Such a variation has to be compared with that achieved in the case of full accretion models, where the variation of $T_\rmn{eff}$ and luminosity is consistently accounted for. Referring to top panel of Fig.~\ref{fig:cmd_acc} (Solar-like mixture), the simplified accretion track is systematically bluer (from 0.01 to 0.04~mag depending on $M_\rmn{p}$) and can reach a maximum difference in magnitude of about 0.1~mag with respect to the full accretion one. On the other hand, the simplified accretion model is much closer to those obtained using the Earth-like mixture (bottom panel). In this case, full and simplified accretion tracks differs in the CMD by less than 0.01~mag in colour and 0.05~mag in magnitude. 

In summary, although the differences between simplified and full accretion models are present also in the CMD, they are quite small when compared to the current observational uncertainties.

\section{Impact on the inferred stellar properties}
\label{sec:rec}
%%
%% Fig: CMD -- recovery Solar Mixture
\begin{figure*}
\centering
\includegraphics[width=\columnwidth]{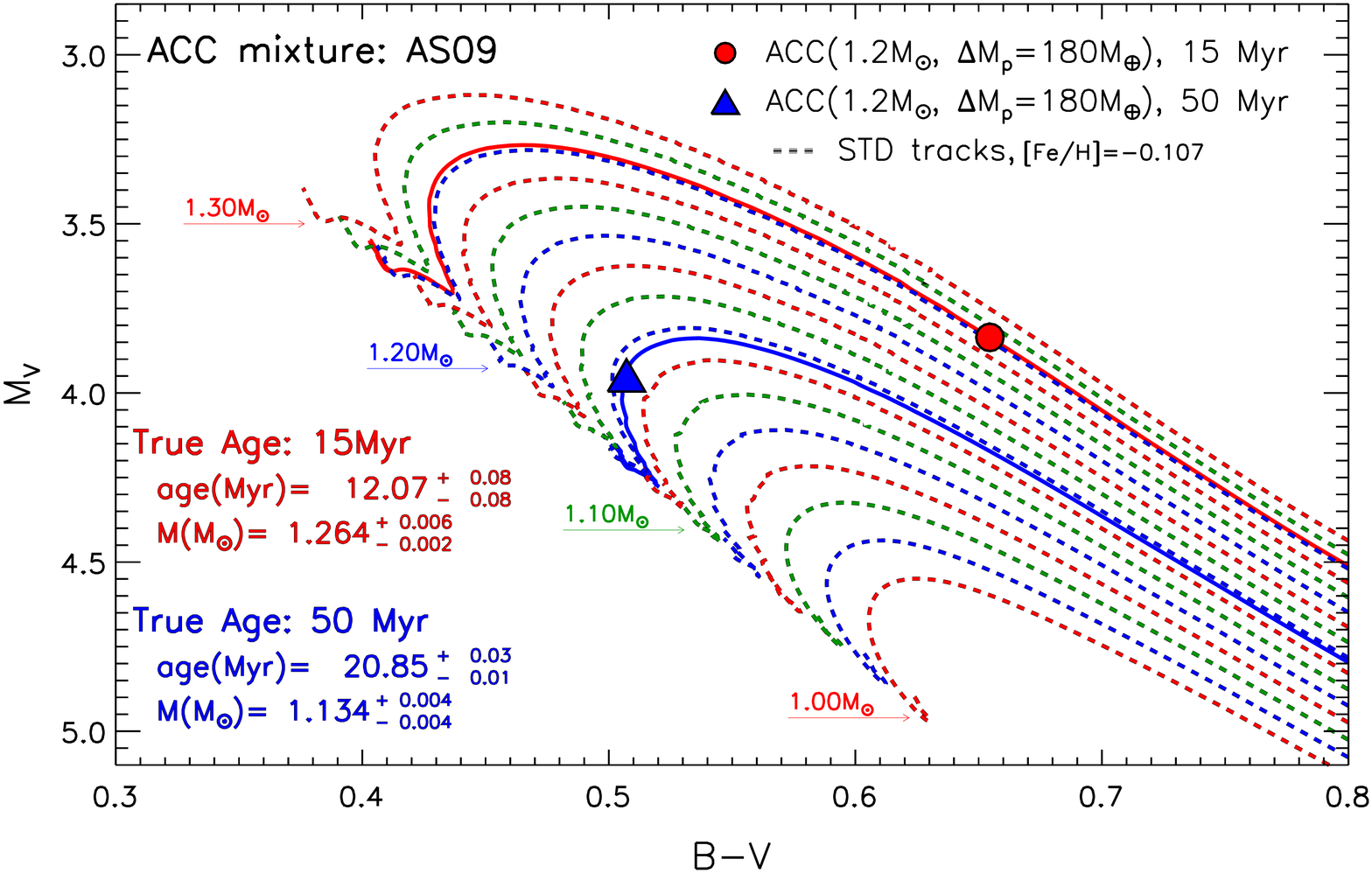}
\includegraphics[width=\columnwidth]{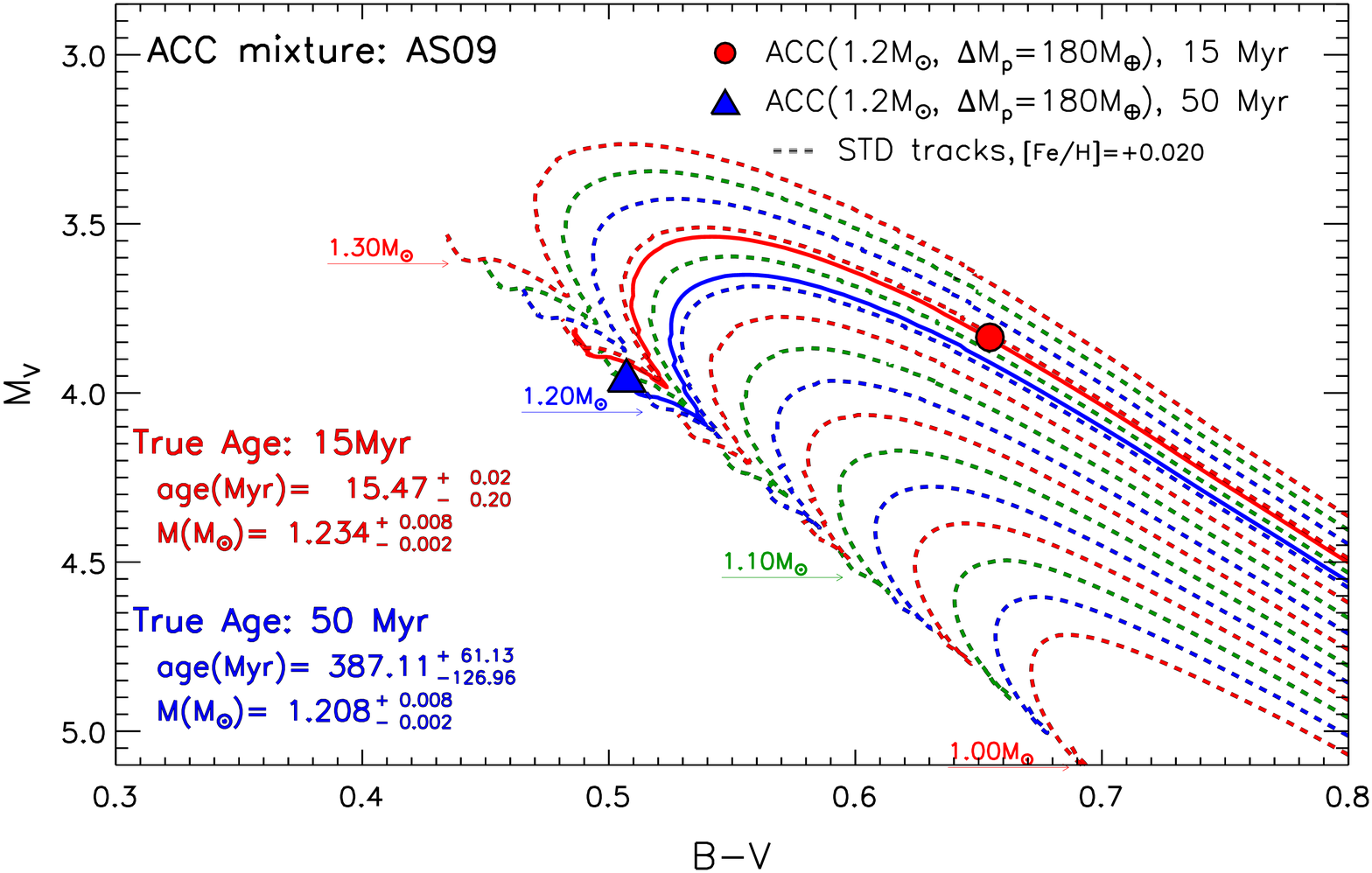}
\caption{Colour-Magnitude diagram with the standard evolutionary tracks used for the recovery (dashed lines). The 15~Myr (filled red circle) and 50~Myr models (filled blue triangle) obtained from the 1.2~\msun{} accreting model with the AS09 Solar-like mixture and $M_\rmn{p}=180$~\mearth{} are shown. The best fit tracks for the 15~Myr (solid red line) and 50~Myr model (solid blue line) are shown too. Left panel: standard models with [Fe/H]$=-0.107$. Right panel: standard models with [Fe/H]$=+0.02$.}
\label{fig:cmd_sinte_mixSun}
\end{figure*}
%%
%% Fig: CMD -- recovery Earth Mixture
\begin{figure*}
\centering
\includegraphics[width=\columnwidth]{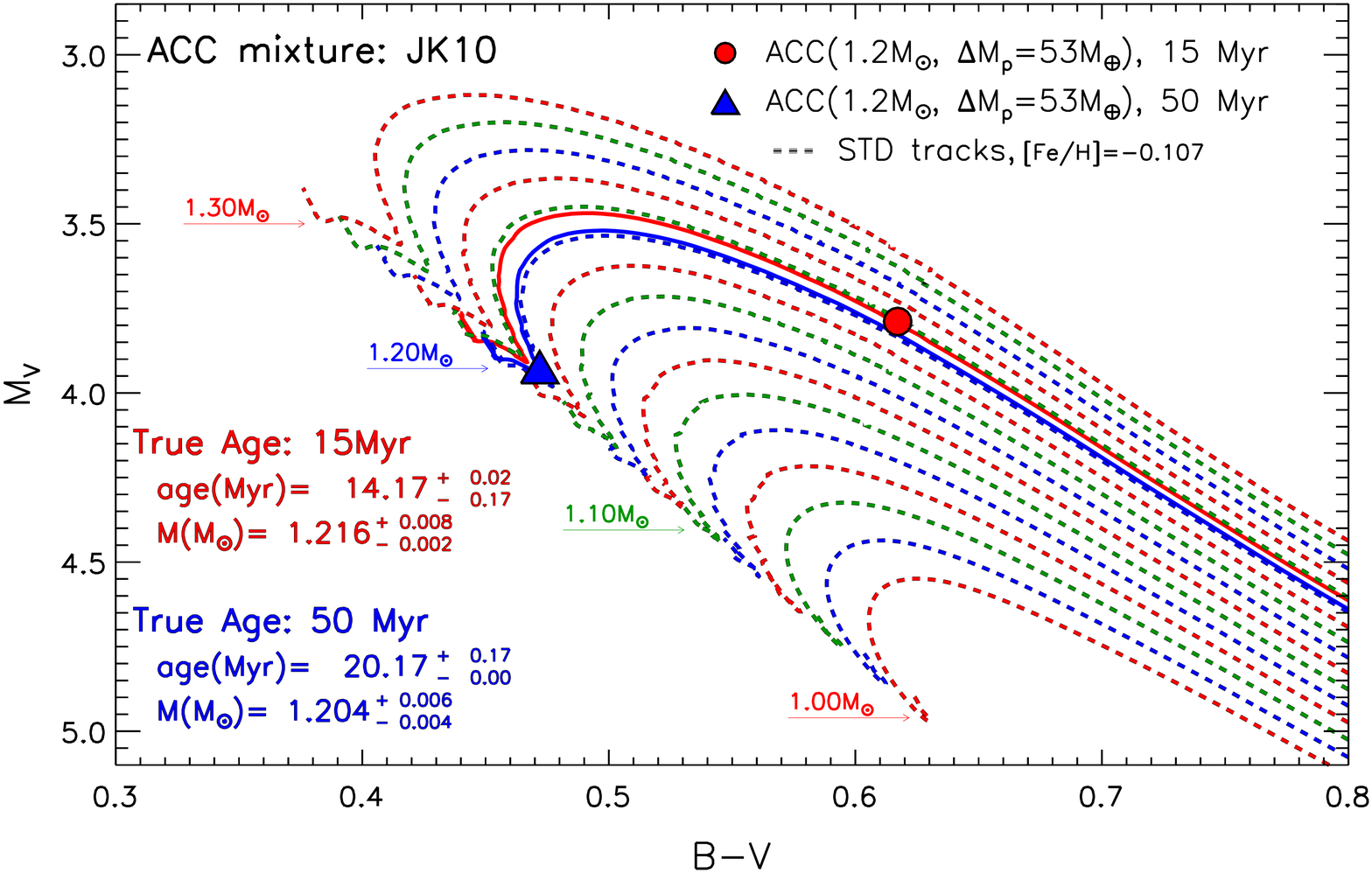}
\includegraphics[width=\columnwidth]{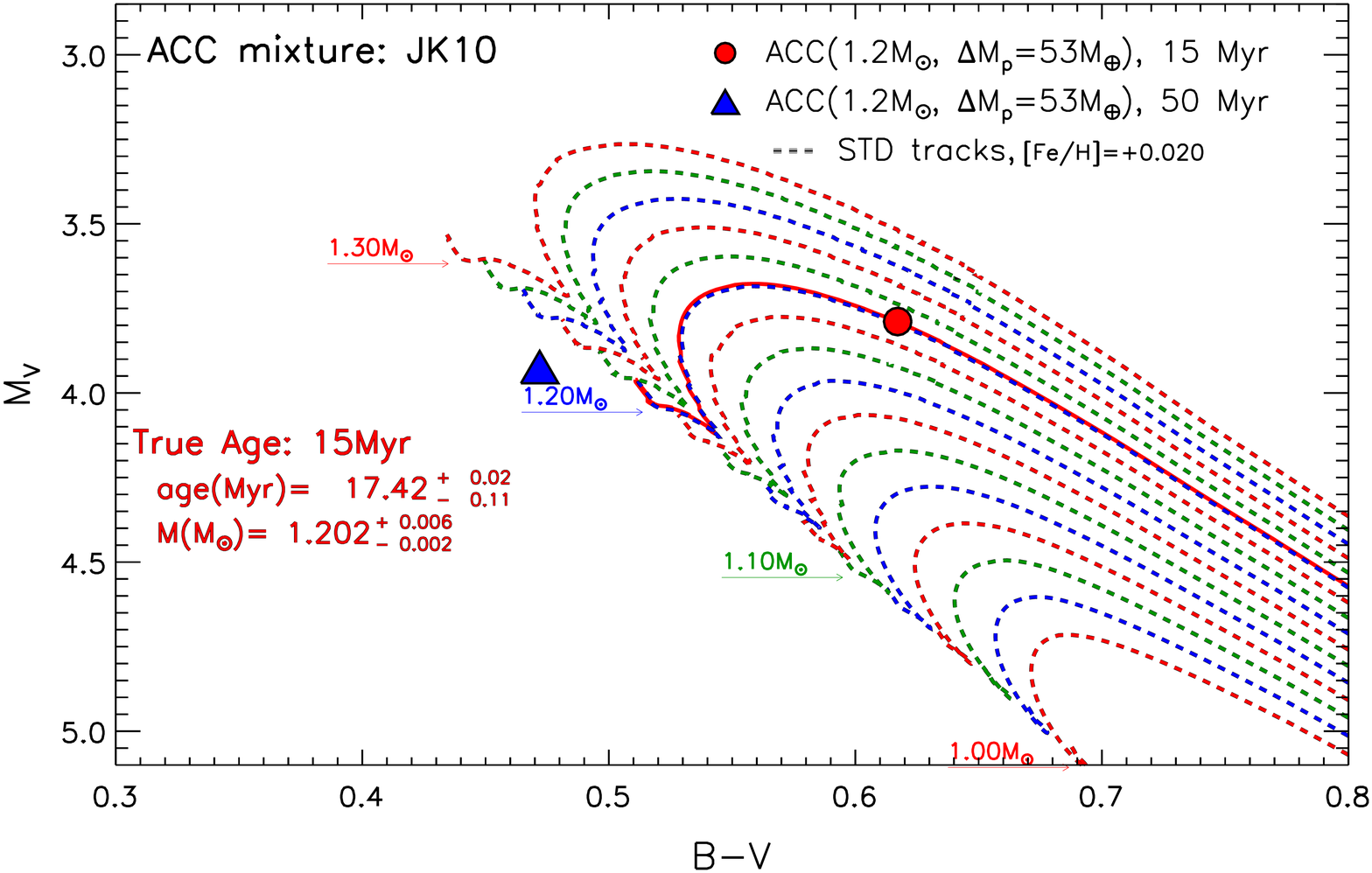}
\caption{The same as in Fig. \ref{fig:cmd_sinte_mixSun} but for models with the JK10 Earth-like mixture in the accreted matter and $M_\rmn{p}=53$~\mearth{}.}
\label{fig:cmd_sinte_mixEarth}
\end{figure*}

Since important stellar quantities such as mass and age are often inferred by comparing theoretical models and observations, the shift in the CMD location of the evolutionary tracks that takes into account the accretion process with respect to the standard ones directly translates into differences in the inferred stellar quantities. In other words, the values of the stellar mass and age (or the mass value of the ingested planet) obtained by means of standard models might be quite different from those based on models which explicitly take into account the accretion process. 

Given such a situation, it is worth to estimate the maximum error in the stellar mass and the age estimated by adopting standard tracks to recover an artificial star generated using accreting ones. We performed this test because usually specific accreting models are not available and the main properties of chemically peculiar stars are derived by means of non accreting stellar evolutionary grids.

Firstly, we sampled synthetic stars with the age of 15~Myr (about the age of Gamma Velorum cluster) and 50~Myr (about the beginning of the \ms) from the accreting evolutionary tracks that show the largest deviation from the standard one. We choose the track of 1.2~\msun{} which follows the ingestion of a $M_\rmn{p}=180~$\mearth{} planet at an age $t_0=10~$Myr with the Solar-like heavy element mixture. We also used the 1.2~\msun{} accreting track that ingested an Earth-like planet of $53~$\mearth{} at $t_0=10~$Myr. We performed the analysis for both the mixtures because of the significant impact of the chemical composition of the ingested planet in the location of the models in the \cmd, as shown in Sect.~\ref{sec:cmd}.

As a second step, we used the standard grid of models with both the low- and high-[Fe/H] to recover the data. To get a good mass resolution, the standard grid has been interpolated at a mass step of 0.001~\msun{}, and each evolutionary track at a time step of $10^4$ yr. The recovery has been performed using a maximum Likelihood technique. We adopted a formalism similar to that described in \citet{gennaro12} to compute the Likelihood (defined in their eq. (3)), the best fit value (the \emph{mode}) and the confidence interval obtained by rejecting the $16$~percent of the total area under the distribution function from the left and the right tail. 

Being interested only to quantify the systematic errors in the inferred age and mass of a star that ingested a planet by using available standard non-accreting models in the recovery procedure, we adopted a very small uncertainty in the synthetic data in the ($B-V$, $M_V$) plane, namely $\sigma_{V}=\sigma_{B} = 0.0005$.

Fig. \ref{fig:cmd_sinte_mixSun} shows the synthetic star in the \cmd{} at an age of 15 and 50~Myr (filled red circle and blue triangle, respectively), extracted from the model of 1.2~\msun{} which follows the accretion of a planet of $M_\rmn{p}=180~$\mearth{} and a solar metal distribution (AS09 mixture). A partial grid  of standard tracks used for the recovery with [Fe/H]$_\rmn{AS09}=-0.107$ (left-hand panel) and [Fe/H]$_\rmn{AS09}=+0.020$ (right-hand panel) is also shown. 

The inferred age and mass of the best fit model depends on the [Fe/H] value of the grid. Using the standard grid with [Fe/H]$_\rmn{AS09}=-0.107$ (left-hand panel), the recovery procedure yields a best estimate of the mass of $1.26$~\msun{} and age of $12$~Myr, while the synthetic star has a true mass of 1.2~\msun{} and an age of 15~Myr. Thus, the inferred mass is over-estimated by about $5$~percent and the age is under-estimated by about $20$~percent. Notice that both the mass and age estimates are not compatible -- within their confidence interval -- with the true ones. The discrepancies between the true and the recovered mass and age increases for the synthetic star at a true age of 50~Myr. In this case, the inferred mass and age are under-estimated, by about $6$~percent and $58$~percent respectively, with a best fit mass of $1.13$~\msun{} and an age of about $21$~Myr. 

If the [Fe/H]$_\rmn{AS09}=+0.020$ standard grid is used in the recovery procedure (right-hand panel) the differences between the true and the estimated values slightly reduce in the case of the artificial star of 15~Myr. The best fit mass is about $3$~percent larger ($1.23$~\msun) than the true one, and the best fit age over-estimated by about $3$~percent. The situation is different in the case of the simulated star at a true age of 50~Myr. Although the inferred stellar mass is over-estimated by less then $1$~percent ($M~=~1.21$~\msun{}) the best fit age is badly recovered and drastically over-estimated ($t~=~388$~Myr). Such an occurrence can be easily understood by remembering that metal richer tracks are cooler and fainter than metal-poorer ones. Hence, the CMD location of the simulated star with a true age of 50~Myr is confused with metal-rich models in a much more advanced phase on the \ms. As well known, this phase is much slower than the \pms{} one and a large variation of the age only marginally affects the luminosity and colour of the star. As a consequence, the age estimate becomes less precise, i.e. the marginalised age distribution gets progressively broader and broader moving towards the \ms, thus causing a broadening of the confidence interval too. Moreover, for the same reason, during the \ms{} phase the models with large ages are denser along the evolutionary track. Thus, their contribution in defining the age distribution is the strongest and this leads to prefer a large best fit age \citep[see also the discussion in][]{gennaro12}. 

We repeated the same analysis for the other two artificial stars of 1.2~\msun{} and true age of 15 and 50~Myr, but extracted from the models which follows the ingestion of an Earth-like planet of $53~$\mearth{}, as shown in Fig.~\ref{fig:cmd_sinte_mixEarth}. 

By adopting the standard grid with [Fe/H]$_\rmn{AS09}=-0.107$ (left-hand panel), the recovery yields -- for the model with the true age of 15~Myr -- a best estimate mass of $1.22$~\msun{} (over-estimated by about $1$percent) and age of $14$~Myr (under-estimated by about $6$~percent). For the synthetic star of a true age of 50~Myr, the estimated mass is well recovered (over-estimated by less than $1$~percent), while the age estimate is much worse, yielding about $20$~Myr (under-estimated by about $60$~percent). If the high-[Fe/H] grid is used to infer the best mass and age (right-hand panel), then the artificial star with a true age of 15~Myr is still recovered quite well, especially the mass which is over-estimated by less than $1$~percent, and with an age of $17$~Myr (over-estimated by $16$~percent). The situation is different if the high-[Fe/H] grid is used to recover the synthetic star of 50~Myr. From the figure it is clearly visible that the simulated data is much bluer than the high-[Fe/H] standard  models and, consequently, the simulated star at 50~Myr is not recovered at all. 

The analysis we performed clearly shows that the systematic bias in the mass and age estimates induced by using standard models which do not take into account the accretion strongly depends on the chemical composition of the ingested planet. Unfortunately, generally no indication about this quantity is available, as observation provides only the actual cluster low-[Fe/H] (assumed to be equal to the initial one) and the high-[Fe/H] (after the accretion) of the chemically peculiar star. 

Nevertheless, the presented simulations also show that it is possible to obtain two limiting cases corresponding to accrete matter with two different heavy elements mixtures, namely the Solar-like (AS09, low-Iron content) and the Earth-like (JK10, high-Iron content). The results described above show that in these two limiting cases, the standard tracks with the high- and the low-[Fe/H] grid of standard models provide a best fit mass not very different from the true one for both the 15 and 50~Myr synthetic stars. On the other hand, the age estimates are worst in particular for stars near the ZAMS. 

Given such a situation, since no a priori information about the chemical composition and mass of the ingested planet is available, the best choice is to use, simultaneously, both the low- and high-[Fe/H] standard tracks to constrain the most plausible range of the mass and age of the peculiar star. 

\section{Conclusions}
We analysed the effect of a single episode of planet ingestion on the evolution of a \pms{} star, to explain the peculiar surface chemical abundance observed in 2MASS~J08095427--4721419 ($\#52$) belonging to the Gamma Velorum cluster. Contrarily to previous investigations, we computed \pms{} models that, for the first time, consistently take into account the accretion effects, i.e. the change of both the stellar structure and the chemical composition. The results of our computations have been compared with those achieved by adopting a simplified accretion method that considers only the effect of the accretion on the surface chemical abundances without any modification of the stellar structure.

The impact of a planet ingestion on the evolution of a star depends on both the characteristics of the star and of the planet. Concerning the star, we analysed the evolution of a model with $M=1.2$~\msun{} with an initial [Fe/H] equal to that of the cluster (i.e. [Fe/H]$^\rmn{cl.}_\rmn{AS09}=-0.107$). This particular value of the stellar mass has been obtained to reproduce the $\#52$ position in the CMD at an age of about 15~Myr (i.e. the mid age value of the Gamma Velorum cluster). Since the surface Iron-abundance enhancement caused by the planet ingestion strongly depends on the convective envelope extension -- which in turn depends on the age -- we focused on three values of the age $t_0$ at which the accretion event is supposed to occur, namely $t_0=10$, 12 and 15~Myr (about the Gamma Velorum cluster age). 

We tuned the ingested planet mass to reproduce the observed [Fe/H]$^{\#52}$ value (i.e. [Fe/H]$^{\#52}_\rmn{AS09}=+0.02$) of the stellar metallicity after the accretion process. For the planet chemical composition, we analysed the adoption of two different heavy elements distributions, namely a Solar-like \citep[][low-Iron content]{asplund09} and an Earth-like one \citep[][high-Iron content]{javoy10}.

We showed that the earlier is the accretion episode (i.e. the smaller is $t_0$), the more extended is the convective envelope, and thus the more diluted is the ingested metal-rich material with the pristine stellar one. Consequently, the mass of the accreted planet required to reproduce the observed [Fe/H]$^{\#52}$ value increases as $t_0$ decreases. We also showed that, at a fixed age $t_0$, the mass of the ingested planet necessary to account for the observed chemical peculiarities decreases by increasing the Iron content in the planet. Indeed, we showed that the needed mass of a planet with the Earth-like chemical composition is smaller than that of a planet with a Solar-like one. For a Solar-like mixture we derived a planet mass $M_\rmn{p}$ of 180, 90 and 7~\mearth{} for an accretion age $t_0$ of respectively, 10, 12 and 15~Myr. The values of $M_\rmn{p}$ reduces to 53, 26 and 2~\mearth{} if an Earth-like mixture is adopted.

The ingestion of metal-rich material affects the stellar structure -- after the accretion episode -- and in particular the extension of the convective envelope with respect to that of standard \pms{} track without accretion. We showed that the mass inside the convective envelope just after the accretion can vary up to 10 percent when the Solar-like mixture and the largest value of the ingested planet (i.e. $M_\rmn{p}=180~$\mearth) are used. Such differences reduces to 4--6~percent by adopting lower $M_\rmn{p}$ values. If the Earth-like mixture is used in the accreted matter then the maximum variation reaches about 4~percent.

The change in the extension of the convective envelope has an impact also on the surface [Fe/H] value after the planet ingestion. Full accretion models predict a surface [Fe/H] value slightly lower than that obtained in the case of a more simplified accretion,  differences that increase by increasing the ingested planet mass. However, the maximum extent of such a difference is always lower than about 0.01~dex, thus negligible if compared to the common observational uncertainty on abundance determination.

We quantified the impact of the accretion on the surface chemical composition. First we analysed the total metallicity $Z^\rmn{new}$ after the planet ingestion. We identified two limiting situation. In the first case, which corresponds to accrete matter with a low-Iron content  (i.e. Solar-like mixture for the planet), the surface metallicity $Z^\rmn{new}$ is significantly increased and more similar to that attained by standard models with the initial peculiar star [Fe/H]$^{\#52}$ (i.e. a maximum metallicity $Z$ increase of about 30 percent). In the second case, which corresponds to accrete a planet with an high-Iron content (Earth-like mixture), the surface metallicity is only marginally enhanced and more similar to that attained by standard non accreting models with the initial cluster [Fe/H]$^\rmn{cl.}$ (a maximum metallicity $Z$ increased of about 10 percent). We also analysed the effect of the accretion on the surface abundances of some elements which have been measured by \citet{spina15} in the peculiar star $\#52$. We showed that both the enrichment patterns we obtained using the Solar-like and the Earth-like mixture for the accreted planet are in agreement with that observed by \citet{spina15}. To better constrain the most suitable mixture for the accreted matter, more precise measurements are needed. 

We analysed the impact of the planet ingestion on the tracks in the Colour-Magnitude Diagram. Depending on the characteristics of the ingested planet, the accretion leads to slightly cooler and fainter models caused by the increase of the metallicity (i.e. radiative opacity) in the external layers. The most visible effects in the CMD occurs when a planet of  $M_\rmn{p}=180~$\mearth{} and the Solar-like mixture is ingested by the star, due to to large metallicity enhancement in the external envelope. In this case variations of about 0.06~mag in colour and about 0.07~mag in magnitude can be achieved. The differences get smaller by reducing the ingested planet mass or by adopting the Earth-like mixture in the accreted matter. The effects of the planet ingestion on the stellar structure and in the CMD are still present long after the accretion, even at the beginning of the \ms{} evolution. 

As generally accreting models consistent with the observed peculiar stars are not available, we investigated the systematic bias in the inferred age and mass due to the use of standard non accreting \pms{} tracks to recover the characteristics of a synthetic star generated using an accreting model. The synthetic star had been obtained fixing the mass to 1.2~\msun{} and selecting two ages, namely 15~Myr (about the estimated age of the Gamma Velorum cluster) and 50~Myr (about the beginning of the \ms). We adopted the accreting model that attains the largest deviation from the standard one (at the same age and mass) for both the Solar-like ($M_\rmn{p}=180$~\mearth) and Earth-like planet ($M_\rmn{p}=53$~\mearth) chemical composition. We performed the recovery using both the low- and high-[Fe/H] standard non accreting models. 

As a general result, while the stellar mass estimate is not severely biased by the adoption of non-accreting stellar models, with a maximum difference between the recovered and the true mass of about 6 percent, the stellar age is badly recovered, especially for synthetic data near the \zams, with a maximum uncertainty of 60 percent. The use of standard models computed for an initial [Fe/H] equal to that of the peculiar star (i.e. [Fe/H]=[Fe/H]$^{\#52}$) provides a quite good estimate of the stellar mass when the ingested planet has a low-Iron content (i.e. Solar-like mixture). On the other hand, the accretion of an Iron-rich planet (Earth-like mixture) is better reproduced by standard track with the pristine initial metallicity (i.e. [Fe/H]=[Fe/H]$^\rmn{cl.}$). 

Thus, standard tracks with both the low- and high-[Fe/H] values can be used to constrain the inferred properties of such chemically peculiar objects when the details of the accretion are not known.

\section*{Acknowledgements}
We would like to thank the reviewer for the useful comments and suggestions that helped us to improve the paper. This work has been supported by PRIN-INAF 2012 (\emph{The M4 core project with Hubble Space Telescope}, PI: L. Bedin), PRIN-INAF 2014 (\emph{The kaleidoscope of stellar populations in globular clusters with Hubble Space Telescope}, PI: S. Cassisi), PRA Universit\`{a} di Pisa 2016 (\emph{Stelle di piccola massa: le pietre miliari dell'archeologia galattica}, PI: S. Degl'Innocenti) and by INFN (Iniziativa specifica TAsP).
%%
%%

%% Bibliography
\bibliographystyle{mn2e}
\bibliography{bibliography}
\label{lastpage}

\end{document}